\documentclass[usenatbib]{mn2e}
\usepackage{natbib}  
\usepackage{epsfig}

\catcode`\@=11
\def\gta{\ifmmode{\,\mathrel{\mathpalette\@versim>\,}}
    \else{$\,\mathrel{\mathpalette\@versim>}\,$}\fi}
\def\lta{\ifmmode{\,\mathrel{\mathpalette\@versim<\,}}
    \else{$\,\mathrel{\mathpalette\@versim<}\,$}\fi}
\def\@versim#1#2{\lower 2.9truept \vbox{\baselineskip 0pt \lineskip
    0.5truept \ialign{$\m@th#1\hfil##\hfil$\crcr#2\crcr\sim\crcr}}}
\catcode`\@=12  
\renewcommand{\[}{\begin{equation}}
\renewcommand{\]}{\end{equation}}
\def\vlos{v_{\rm los}} 
\def\cN{{\cal N}}\def\cS{{\cal S}}

\def\pr{{\rm Pr}_{\rm o}}

\def\p{\partial}
\let\boldgrk=\gkvecten
\let\boldgrksc=\gkvecseven

\def\gkthing#1{{\mathchoice%
	{\hbox{{\boldgrk\char#1}}}
	{\hbox{{\boldgrk\char#1}}}
	{\hbox{{\boldgrksc\char#1}}}
	{\hbox{{\boldgrksc\char#1}}}}}

\def\vtheta{\gkthing{18}}

\def\vmu{\gkthing{22}}

\def\vsigma{\gkthing{27}}

{\newif\ifnotend
\notendtrue
\def\veclist{ABCDEFGHIJKLMNOPQRSTUVWXYZabcdefghijklmnopqrstuvwxyz.}
\def\top#1#2.{#1}
\def\tail#1#2.{#2.}
\loop\expandafter\xdef\csname v\expandafter\top\veclist\endcsname%
{{\noexpand\bf\expandafter\top\veclist}}
\edef\veclist{\expandafter\tail\veclist}
\if\veclist.\notendfalse\fi\ifnotend\repeat}
 \def\s{{\rm s}}
\def\df{{\sc df}}\def\pdf{{\sc pdf}}
\def\mas{\,{\rm mas}}

\def\Rc{R_{\rm c}}

\def\kms{\,{\rm km}\,{\rm s}^{-1}}

\def\yr{\,{\rm yr}}

\def\pc{\,{\rm pc}}
\def\kpc{\,{\rm kpc}}

\def\e{{\rm e}}
\def\d{{\rm d}}

\def\feh{\hbox{[Fe/H]}}

\def\afe{[\alpha/\hbox{Fe}]}
\def\dex{\,{\rm dex}}
\def\figref#1{Fig.~\ref{#1}}
\newcommand{\beq}{\begin{equation}}
\newcommand{\eeq}{\end{equation}}

\title[Analysing surveys of our Galaxy]
{Analysing surveys of our Galaxy I:\\ basic astrometric data}

\author[P. McMillan and J. Binney]{Paul J. McMillan$^1$\thanks{E-mail:
p.mcmillan1@physics.ox.ac.uk} and James
Binney$^1$\\
$^{1}$ Rudolf Peierls Centre for Theoretical Physics, Keble Road, Oxford OX1 3NP, UK\\
}
\begin{document}

\date{September 20, 2011}

\pagerange{\pageref{firstpage}--\pageref{lastpage}} \pubyear{2011}

\maketitle

\label{firstpage}

\begin{abstract}
We consider what is the best way to extract science from large surveys of the
Milky Way galaxy. The diversity of data gathered in these surveys, together
with our position within the Galaxy, imply that science must be extracted by
fitting dynamical models to the data in the space of the observables. Models
based on orbital tori promise to be superior for this task than traditional
types of models, such as N-body models and Schwarzschild models. A formalism
that allows such models to be fitted to data is developed and tested on
pseudodata of varying richness drawn from axisymmetric disc models. 
\end{abstract}


\section{Introduction}

A major thread of current research is work directed at understanding the
origin of galaxies. There are excellent prospects of achieving this goal by
combining endeavours in three distinct areas: observations of galaxy
formation taking place at high redshift, numerical simulations of the
gravitational aggregation of dark matter and baryons, and studies of the
Milky Way. The latter field is dominated by a series of major observational
programs that started fifteen years ago with ESA's Hipparcos mission, which
returned parallaxes and proper motions for $\sim10^5$ stars
\citep{Hipparcos}.  Hipparcos established a more secure astrometric reference
frame, and the UCAC-3 catalogue \citep{UCAC} uses this frame to give proper
motions for several million stars. These enhancements of our astrometric
database have been matched by the release of major photometric catalogues
DENIS \citep{Denis}, 2MASS \citep{2MASS}, SDSS \citep{Sloan}, SEGUE
\citep{Segue}, and the accumulation of enormous numbers of stellar spectra,
starting with the Geneva--Copenhagen survey \citep[][hereafter
GCS]{Nordstrom04,HolmbergNA} and continuing with the SDSS, SEGUE and RAVE
\citep{Steinmetz,DR3} surveys. Major programs to obtain low-dispersion stellar
spectra are currently underway -- on completion the SEGUE and RAVE surveys
will each provide $\sim 5\times10^5$ spectra.  These spectra
yield good line-of-sight velocities and estimates of $\feh$ errors of
$\sim0.1\dex$ and a coarse estimate of $\afe$.  Three surveys
(APOGEE, ESO-Gaia and HERMES) are currently being being prepared that will
obtain large numbers of spectra with resolutions in the range $20\,000 -
40\,000$ from which abundances of significant numbers of elements can be
determined. Several important photometric surveys are currently under way,
including the PanSTARRS survey \citep{PanSTARRS} which is obtaining $griz$ photometry
through a large part of the sky, and surveys of the bulge region and the
Galactic plane in the near-IR with the VISTA telescope. The era of great
Galactic surveys will culminate in ESA's Gaia mission, which is scheduled for
launch in early 2013 and aims to return photometric and astrometric data for
$10^9$ stars and low-dispersion spectra for $\sim10^8$ stars. 

The Galaxy is an inherently complex object, and the task of interpreting
observations is made yet more difficult by our location within it.
Consequently, the ambitious goals that the community has set itself, of
mapping the Galaxy's dark-matter content and unravelling how the Galaxy was
assembled, can probably only be attained by mapping observational data onto
sophisticated models. This paper is the first in a series in which we use
a new method of analysing dynamical models to interpret data 
from surveys of different
types.  Here we introduce and test the basic principles using mock survey
catalogues of the region $b>30^\circ$ that contain data of Gaia-like quality
that progresses in completeness from photometry plus proper motions, through
photometry, proper motions, parallaxes and line-of-sight velocities.  In
subsequent papers we will extend the formalism to include spectrophotometric
data and apply it to real catalogues.

The paper is organised as follows. Section \ref{sec:why} explains the
fundamental importance of equilibrium dynamical models for the problem in
hand. Section
\ref{sec:type} discusses the feasibility of using N-body models to interpret
surveys of the Milky Way and outlines our preferred strategy. Section
\ref{sec:strategy} presents the formulae used to calculate the likelihood of
a catalogue given a model, while in Section \ref{sec:optimise} we explain how
we find the optimum values of the model's parameters and their statistical
uncertainties. Section \ref{sec:test} describes how we construct a
pseudo-catalogue from a model and a number of tests of our method that we have
run using a Gaia-like pseudo-catalogue.
Section \ref{sec:discuss} outlines a number of directions for further work and
future developments of our methodology. Section \ref{sec:conclude} sums up.

\section{Why we need Galaxy models}\label{sec:why}

Models of the Galaxy have a key role to play for several reasons. First they
enable one to understand and compensate for observational biases, which
dominate all data sets on account of our location in the Galaxy's dust- and
gas-rich disc. Second they provide the natural means of tying together data
from different surveys -- surveys may concentrate on obtaining photometric,
astrometric or spectral data and different surveys probe magnitude ranges and
therefore constrain the Galaxy over complementary distance ranges. A model
can assemble this complementary information into a single coherent picture. 

In the 1980s Bahcall and his collaborators moved Galactic astronomy an
important step forward by introducing models of the stellar content of the
Milky Way that were inspired by observations of external galaxies
\citep[][and references therein]{BahcallS,Ratnatunga}. Although these models included the kinematics of stars,
velocities were not required to be consistent with Newton's laws of motion.
In the Besan\c con model \citep{Robin03}, Newton's laws are used to connect
the vertical structure of the disc to the distribution of vertical
velocities, $W$, but the large-scale structure of the Besan\c con model of
the Galaxy is not constrained by Newton's laws. The most recent  incarnation
of this type of model is presented by \cite{Sharma11}.

Fully dynamical models, that is models in which the distribution of stars is
consistent with Newton's laws of motion, are now required for several
reasons. Most obviously, the distribution of dark matter can be deduced from
observations of stars and gas only by assuming that the Galaxy is in an
approximate steady state, and interpreting observations in the framework of an
{\it equilibrium\/} dynamical model -- if we do not assume a steady state (which
strictly speaking must be a false assumption), any mass distribution is
consistent with any phase-space distribution of the baryonic matter.
Inferences about the mass content can only be drawn because
a sufficiently large central concentration of mass would imply a rapid
collapse of the baryonic component from its observed configuration, and a
sufficiently small mass concentration would imply that the baryonic component
would fly apart. We must deduce the actual mass distribution by assuming
that  the baryonic component is in an approximate steady state,
thus ruling out large-scale contraction or expansion. Hence equilibrium
dynamical models are fundamental for achieving a major goal of
contemporary astronomy.

An attractive feature of equilibrium dynamical models is that they reduce the
Galaxy from a six-dimensional to a three-dimensional object: without the
assumption of dynamical equilibrium, we have to specify the distribution of
stars in six-dimensional phase space, while the assumption of equilibrium
together with Jeans' theorem makes it sufficient to specify the distribution
of stars in three-dimensional space of isolating integrals. For obvious
reasons, objects can be imagined in three dimensions much more easily
than they can be imagined in four or higher dimensions, so the reduction in
dimensionality from six to three is a major simplification. This reduction
also vastly reduces the amount of information required to specify a model: if
we measure each position and velocity with a precision of, say, 1 percent, a
six-dimensional model with less than $10^{12}$ resolution elements will
degrade the resolution of the raw data, while a three-dimensional model
requires only $10^6$ resolution elements for a faithful representation of the
data.

A dynamical model connects objects that we can observe to objects that we
have not observed, either because they are distant and therefore faint, or
because they are obscured by dust. For example \cite{MayB} showed that if the
stellar halo were in virial equilibrium, more than half the stars of the
stellar halo would be on orbits that bring them through the solar
neighbourhood, so in principle from measurements made in a small volume
around the Sun we could determine the density of stars on more than half the
halo's populated orbits. 

The Galaxy is certainly not in a steady state. Most obviously because it has
a bar inside $\sim3\kpc$ and spiral arms in the surrounding disc. These
features not only rotate with various pattern speeds, but on longer
timescales change their structure and may well decay.  Moreover, by
scattering stars and dark-matter particles they drive secular evolution of
all the Galaxy's components. The resulting secular evolution of the disc has
been studied for half a century
\citep[e.g.][]{Roman54,SpitzerS,CarlbergS,BinneyL,DehnenB,SchoenrichBII} and
has clearly played a major role in determining the observed state of the
Galaxy. Secular evolution is most readily modelled by adding perturbations to
the Hamiltonian of an equilibrium model, that cause the model to move through
a series of equilibrium states. Hence equilibrium models are the key to
modelling secular evolution as well as to determining the distribution of
dark matter.

Data from the SDSS survey revealed that a significant proportion of the
stellar halo has yet to phase mix properly, and that it contains numerous
tidal streams \citep{Bell}. These streams have considerable potential for
mapping the Galaxy's gravitational field and  distribution of dark matter.
Galaxy models of the type we advocate in this paper would seem to offer the
best hope of fully exploiting the potential of tidal streams \citep{EyreB11}.

The complexity of the Galaxy is such that we cannot realistically hope
eventually to build a dynamical model that accounts for {\it every\/}
observation in detail. For example, it is known that stellar discs are highly
responsive objects, and much of the Galaxy's current spiral structure will
represent an ephemeral response to noise driven by star formation and
fluctuations in the density of the dark-matter halo.  The Galaxy's warp is
likely to be driven by such density fluctuations. In these circumstances we
should aim for a sequence of dynamical models of increasing complexity and
realism. We aim first for an axisymmetric, equilibrium model and identify the
most prominent features in the data that cannot be explained by such a model.
Then we aim for a steady-state barred model and see how much more of the data
such a model can explain. Next we include either the warp or spiral structure
by perturbing our best barred model, and see how much better we can fit the
data. By proceeding in this way we can hope to reach the point at which we
feel we have a model that provides as good a fit to the data as it is
reasonable to expect, and we will have learnt along the way a great deal
about the Galaxy's contents and manner of operation. In this paper we are
concerned with the first stage in this journey, the construction of
axisymmetric models.

\section{Types of Galaxy model}\label{sec:type}

\subsection{N-body models}

The simplest galaxy models to construct are N-body models. The initial
conditions from which an N-body model is integrated are generally not
consistent with an equilibrium model because such initial conditions can only
be chosen if one is already in possession of the desired model. Therefore the
equations of motion must be integrated for some time to allow the model to
settle to an equilibrium through relaxation.  This period of
relaxation has two unfortunate consequences.  First, equilibrium is reached
only asymptotically in time, and in the case of a disc galaxy significant
relaxation may continue for an inconveniently long time.  Second, it is not
clear what initial conditions are required to achieve a given configuration.
Recently the ``made-to-measure'' (M2M) technique introduced by \cite{SyerT}
has been sharpened by \cite{deLorenzi}, \cite{Dehnen09} and others into an
effective way to refashion a model that somewhat resembles the target galaxy
into a model that represents that galaxy to good accuracy. Consequently the
dynamical models that currently best approximate the Galaxy are M2M models
\citep{Bissantz04,Martinez-VG}.

While M2M models can be extremely useful, they are less than ideal from a
number of perspectives. First, an M2M model is specified by $N$ phase-space
coordinates and particle weights, where $N\gta10^5$ is a large number. This
specification is at once cumbersome and non-unique; at a later time the
coordinates will be different while the model will be the same. An equally
good model of the same galaxy made by a different group will use a different
mix of orbits so the weights will be different and it will not be evident
that the two models are equivalent. Second, when the model's phase space has
significant stochastic regions, it is not possible to halt dynamical
evolution of the model during the period that orbits are followed in order to
optimise their weights. Finally, there is a fundamental problem
with any particle-based model of the Galaxy that was pointed out by
\cite{BrownVA}: important information about the Galaxy is contained in
observations of low-luminosity stars that can only be observed close to the Sun, so
such stars must be represented in any complete Galaxy model. However,
orbiting particles that are at one moment near the Sun are some time later
far from the Sun. If particles are treated as stars, those that represent
low-luminosity stars will rarely contribute to observables because the particle will
usually to be too far from the Sun to be visible. Consequently, the number of
particles contributing to observables will be much smaller than the number of
particles in the model.  One might hope to circumvent this problem by
considering particles to be representatives of a stellar population that has
a well defined luminosity function. Then when a particle is far from the Sun,
one could be sure to draw a representative star luminous enough to be visible
from the Sun. But then for consistency {\it many} stars must be drawn from
the particle when it is near the Sun. Clearly the model's shot noise will be
adversely affected if each nearby particle contributes to the observables a
host of stars with exactly the same phase-space coordinates, just as it will
be if the number of stars is much less than the number of particles.

To escape from this sampling problem we need to be able to sample the solar
neighbourhood much more densely than far-flung parts of the Galaxy. That is,
our model must yield the probability density of particles rather than a
specific realisation of this density -- we need to know the phase-space
density $f$ of stars, for armed with
$f$ we can populate the solar neighbourhood with large numbers of
mostly low-luminosity stars and remote parts of the Galaxy with smaller numbers of
exclusively luminous stars.

In principle it {\it is\/} possible to determine the phase-space density $f$
from an N-body model because Liouville's theorem assures us that phase-space
density is constant along orbits, so the phase-space density is known at the
location of every particle provided the initial conditions sampled a well
defined sampling density $f_{\rm s}$ \citep[e.g.\ \S4.7.1 of][]{BT08}. Hence
the phase-space density at any point in the phase space of the final model
can be estimated by six-dimensional interpolation. Sampling the phase-space
density obtained in this way is not easy, but is achievable with the
Metropolis algorithm for example \citep[e.g.][ \S4.3]{TCP}. However, these
steps are highly non-trivial and to the best of our knowledge no N-body model
has been successfully resampled.

The problem of determining the distribution function (\df) $f$ is made more
challenging by the fact that we need not one \df\ but a \df\  $f_\alpha$ for each
stellar population $\alpha$: as a minimum the model must predict the spatial
distribution and kinematics of stars that lie in each of several regions of
the $(\feh,\afe)$ plane \citep[e.g.][]{SchoenrichBII}. It would not be
straightforward to adapt an N-body model so that it yielded several \df s
simultaneously, although ab initio models of galaxy formation can assign
stellar parameters to particles \citep[e.g.][and references
therein]{Simon+11}.

\subsection{Torus models}

The modelling technique introduced by \cite{SchwarzI} has become the standard
tool for analysing the dynamics of early-type galaxies
\citep[e.g][]{Krajnovic}, especially in connection with searches for central
black holes \citep[e.g.][]{Gebhardt03}. This technique involves first
integrating a number of orbits in the adopted gravitational potential and
then seeking weights for these orbits such that the weighted sum of the
orbits reproduces the observational data to acceptable precision.  

Torus modelling is analogous to Schwarzschild modelling except that orbits,
which are essentially time series of phase-space points, are replaced by
orbital tori, which are analytic expressions for the three-dimensional
continuum of phase-space points that are accessible to a star on a given
orbit. Whereas an orbit is labelled by its initial conditions, a torus is
labelled by its three action integrals $J_i$; whereas position on an orbit is
determined by the time $t$ elapsed since the initial condition, position on a
torus is determined by the values of three angle variables $\theta_i$, one
canonically conjugate to each action $J_i$. For a list of advantages arising
from the replacement of orbits by tori, see \cite{BinneyM11}, and for a
summary of how orbital tori are constructed and references to the papers in
which torus dynamics was developed, see \cite{McMillanB08}. Note that torus
models provide the ideal framework within which to study features such as
spiral arms or warps because they come with angle-action variables; these
coordinates were invented for perturbative studies of the solar system and,
as \cite{Kaasalainen94} showed, perform spectacularly well when they have been
derived by constructing tori.

Although it is possible to treat the weights of tori as independent unknowns
to be fitted to the data just as in Schwarzschild modelling, a better
procedure is to derive the weights from the model's \df. By Jeans' theorem,
the \df\ of a steady-state model can be taken to be a function $f(\vJ)$ of
the action integrals. Following Binney (2010; hereafter B10) we make $f$ an
analytic function of certain parameters $a_i$ in addition to $\vJ$, and we
fit the model to the data by varying the $a_i$.  

The impact of shot noise on the model is minimised if all tori have the same
weight, and this will be the case if the density of used tori in action space
samples the \df. We endeavour to ensure that this condition is met, at least
to a good approximation.

\section{Assessing a model}\label{sec:strategy}

Our modelling strategy
is as follows. For each trial gravitational potential we construct a library
of orbital tori that has a known sampling density in action space. For each
of the Galaxy's identified stellar populations (such as G-dwarfs or K-giants
of given metallicity) we have a luminosity function $F(M)$ and a trial
\df\ with parameters that have to be chosen for that population. For given
values of these parameters, we can infer the weight of each of the library's
tori from the \df. Given these weights and the numerically-determined mapping
from actions and angles to Cartesian phase-space variables, we can find the
probability of finding a star of a given absolute magnitude at any point
($\vx,\vv$) in phase space, and therefore at any point in the space of
observables. Hence for given values of the parameters we can evaluate the
likelihood of the data given the model that is defined by the current
gravitational potential and the current parameters. We find the values of the
parameters that maximise the likelihood for the given potential, and then
repeat the process for a different potential. In this way we determine what
range of gravitational potentials and \df s is consistent with the data.

The Galaxy's gravitational potential $\Phi$ is generated by the sum of the
mass densities of {\it all\/} the Galaxy's populations, including that of its
dark-matter particles. Our knowledge of $\Phi$ remains quite limited, so in
the coming years the challenge is to use the kinematics of stars and gas to
constrain $\Phi$ more tightly. Then Poisson's equation can be used to derive
the dark-matter density as the difference between the density that generates
$\Phi$ and the density of stars and gas.  Once the distribution of dark
matter is fairly well known, it will be appropriate to seek a \df\ for the
dark matter and construct a completely self-consistent model of the Galaxy.
At the present stage of our understanding a concern for self-consistency
would be premature because many \df s for dark matter will be consistent with any
plausible density distribution of dark matter. Consequently, nothing is to be gained
by specifying the dark matter's \df\ until there are kinematic constraints on
it, presumably provided  by experiments that detect dark-matter
particles underground.

In this paper our focus is on the determination of the \df\ of stars
given typical observational data and  a trial form for $\Phi$. We defer to a
subsequent paper a study of the effects of changing the assumed form of
$\Phi$ and thus the extent to which given data allow one to constrain $\Phi$.

\subsection{Basic notation}\label{sec:basic}

We now lay out the formulae that are required to implement the strategy just
described. The data consist of a catalogue that for $N$ stars gives accurate
values of the Galactic coordinates $(b,l)$, data such as apparent magnitude
$m$, colour $V-I$, and line-of-sight velocity $\vlos$ that have moderate
errors, and values of the parallax $\varpi$, proper motion $\vmu$, surface
gravity $g$ and metallicity $Z$ that are probably significantly in error. We
group the variables into two sets, the basic variables
 \[\label{eq:defsu}
\vu\equiv(b,l,m,\varpi,\vmu,\vlos)
\]
 and additional astrophysical variables
 \[\label{eq:s}
\vs\equiv (V-I,g,Z).
\]
 Note that $\vu$ has seven components, effectively a star's phase-space
coordinates $(\vx,\vv)$ and its apparent magnitude $m$. For now we neglect
interstellar extinction. Then the star's absolute magnitude $M$ is
effectively specified by $\vu$ because the star's
distance is fixed by $\vx$. 

We assume that the errors in the observed
quantities are independent and can be modelled by Gaussian probability
distributions
\[
G(u,\overline{u},\sigma)\equiv{\e^{-(u-\overline{u})^2/2\sigma^2}\over\sqrt{2\pi\sigma^2}}.
\] 
 We attach primes to the true values of measured quantities to distinguish
them from the measured values, so we generically have that the probability of
measuring a value $u$ is
 \[\label{eq:Gauss}
\pr(u)=\int\d u'\,G(u,u',\sigma)\,{\rm Pr}_{\rm model}(u'),
\]
Any quantity such as the parallax that is not given in the catalogue can be
considered to have a sufficiently large $\sigma$ that the Gaussian
density is effectively constant for all relevant values of the variable.

For brevity we use the notation
\[
G_i^j(\vu,\vu',\vsigma)\equiv\prod_{k=i}^jG(u_k,u_k',\sigma_k).
\]

In this paper we restrict ourselves to the case of a single stellar
population. This assumption ensures that there are no correlations
between stellar type and kinematics: the distribution of stars in phase space
is independent of their luminosities, colours, metallicities, etc. In this
case  we  can confine discussion to the components of $\vu$ and neglect $\vs$.
We assume that the luminosity function $F(M)$ satisfies the normalisation
 condition
 \[
1=\int_{-\infty}^\infty\d M\,F(M)
\]
and that the \df\ $f(\vJ)$ is normalised such that 
\begin{equation}
1 = \int_{\mathrm{all\ } \vJ,\vtheta}\d^3\vJ \,\d^3\vtheta\, f(\vJ).
\end{equation}
 Note that $(\vJ,\vtheta)$ determines $(\vx,\vv)$, so the set
$(M,\vJ,\vtheta)$ fixes the true values of a star's observables, $\vu'$.

We require the probability of observing a star with observables in $\d^7\vu'$
given that the star satisfies the selection criteria of the survey. If
$\d P(\vu')$ is the probability that a  star chosen randomly from the Galaxy
as a whole lies in $\d^7\vu'$ around $\vu'$, then
 \[
\d P(\vu')=\d P(\vu'|\hbox{in survey})P(\hbox{in survey}).
\]
 But
 \[
\d P(\vu')=f(\vJ)F(M){\p(M,J,\vtheta)\over\p(\vu')}\,\d^7\vu',
\]
 so
\[\label{eq:noerror}
P(\vu'|\hbox{in survey})\,\d^7\vu'={f(\vJ)F(M)
\over P(\hbox{in survey})}{\p(M,J,\vtheta)\over\p(\vu')}\,\d^7\vu'.
\]
 To take into account observational errors, we fold the probability
distribution (\ref{eq:noerror}) with the Gaussian kernel and have that the
probability density of stars in the space of observables that is predicted by
the \df\ $f$ is
 \begin{eqnarray}\label{eq:prelimP}
\pr(\vu|f)&=&\int\d^7\vu'\,G_1^7(\vu,\vu',\vsigma)
P(\vu'|\hbox{in survey})\nonumber\\
&=&{\int\d^3\vJ\,f(\vJ)\int\d^3\vtheta\int\d M\,
F(M)G(\vu,\vu',\vsigma)\over P(\hbox{in survey})},
\end{eqnarray}
 where $\vu'(M,\vJ,\vtheta)$ is the vector of observables corresponding to
the given absolute magnitude and phase-space position.  

 Since $\pr(\vu|f)$ is a properly normalised probability density function 
(pdf), we have
\begin{eqnarray}\label{eq:Psurvey1}
P(\hbox{in survey})&=&\int\d^3\vJ\,f(\vJ)\int\d^3\vtheta\int\d M\,
F(M)\cr
&&\times \int_{\rm S}\d^7\vu\,G(\vu,\vu',\vsigma)
\end{eqnarray}
 where the subscript S on the second integral implies integration through the
range of observables encompassed by the survey. Usually, there is no
restriction on the velocities, so the Gaussian factors in $\mu_b$, etc,
integrate out to unity. Similarly the parallax will not be restricted a
priori, so its Gaussian factor will integrate to unity. We assume that the
Gaussian factors in $b$ and $l$ integrate to either zero or unity, depending
on whether the true values $b'$ and $l'$ lie in the surveyed region because
the observational errors in the sky coordinates are so small. Hence only the
integral over apparent magnitude $m$ need be considered, and in this paper we
make the assumption that the errors in $m$ are small compared to the width of
the luminosity function, so we assume that the remaining integral is zero or
unity depending on whether the true apparent magnitude $m'$ lies within the
survey limits. Thus we adopt
 \begin{equation}
P(\hbox{in survey})=\int\d^3\vJ\,f(\vJ)\int_{\rm S}\d^3\vtheta\int_{-\infty}^{M_{\rm
crit}(r)}\hskip-10pt\d
M\,F(M),
\end{equation}
 where $r(\vJ,\vtheta)$ is the heliocentric distance of the specified point
in phase space and
 \[\label{eq:defsMcrit}
M_{\rm crit}(r)=m_{\rm lim}-5\log_{10}(r/10\pc)
\]
 with $m_{\rm lim}$ the limiting apparent magnitude of the survey. We define
the survey's selection function to be
 \[
\phi(\vJ)=\int_{\rm S}\d^3\vtheta\int_{-\infty}^{M_{\rm crit}(r)}\hskip-10pt\d
M\,F(M),
\]
 Then we have
 \[\label{eq:Psurvey2}
P(\hbox{in survey})=\int\d^3\vJ\,f(\vJ)\phi(\vJ),
\]
 and (\ref{eq:prelimP}) can be written
 \begin{equation}\label{eq:finalP}
\pr(\vu|f)={\int\d^3\vJ\,f(\vJ)\int_{\rm S}\d^3\vtheta\int\d M\,
F(M)G(\vu,\vu',\vsigma)\over\int\d^3\vJ\,f(\vJ)\phi(\vJ)}.
\end{equation}
From these individual
probabilities we construct the log likelihood of a survey for a given model: 
 \[\label{eq:trial}
L\equiv\sum_\alpha\ln\left[\pr(\vu_\alpha|f)\right],
\] 
 where the sum is over the stars in the survey.  In an appendix we
demonstrate explicitly that $L$ is stationary when the trial \df\ used to
evaluate it is equal to the \df\ that was used to produce the catalogue being
analysed.

The integrals over $\vJ$ in equation (\ref{eq:finalP}) are most conveniently
done by Monte-Carlo integration.  We arrange that the points at which the
integrand is evaluated  sample the \df , so we may replace
$\int\d^3\vJ\,f(\vJ)$ by $N^{-1}\sum_k$:
 \[\label{eq:Pru}
\pr(\vu|f)={\sum_{k=1}^N
\int_{\rm S}\d^3\vtheta\int
\d M\,F(M)G_1^7(\vu,\vu'_k,\vsigma)\over\sum_k\phi(\vJ_k)}
\]
 where $\vu'_k(M,\vJ_k,\vtheta)$.

The evaluation of the integral in equation (\ref{eq:Pru}) can be simplified
by taking advantage of the fact that the errors in $(b,l)$ are small so the
integrand is non-negligible only when $(b',l')$ lies close to $(b,l)$. We can
approximate the Gaussians in these variables as $\delta$-functions and
integrate them out analytically: doing so we introduce a Jacobian because
 \begin{eqnarray}
\int\d^3\vtheta&&\!\!\!\delta(b-b')\delta(l-l')\nonumber\\
&=&
\int\d b'\d l'\d\varpi'\,\left|{\p(\vtheta)\over\p(b',l',\varpi')}\right|
\delta(b-b')\delta(l-l')\nonumber\\
&=&\int\d\varpi'\,\left|{\p(\vtheta)\over\p(b,l,\varpi')}\right|.
\end{eqnarray}
 The evaluation of the  Jacobian is described in the Appendix of 
\cite{BinneyM11}. Now we have
 \begin{equation}\label{eq:short}
\pr(\vu|f)
={\sum_k\!\int\!\d r'\left|{\p(\vtheta)\over\p(b,l,r')}\right|
\int\d M\,F(M)G_3^7(\vu,\vu'_k,\vsigma)\over\sum_k\phi(\vJ_k)},
\end{equation}
 where the integration over the true parallax $\varpi'=1/r'$ has been
transformed into an integration over true heliocentric distance. The integral
over $M$ simply returns the fraction of the stellar population that is
consistent with the measured apparent magnitude at distance $r'$.  We use
equation (\ref{eq:short}) for $\pr(\vu|f)$ in equation (\ref{eq:trial}) when
evaluating the likelihood.

We determine $\phi(\vJ)$ as follows. For each value of $\vJ$ we choose at
random a point $\vtheta_i$ in angle space. The point corresponds to a
distance $r$ and sky coordinates $(b,l)$.  If these coordinates lie in the
survey volume, stars with brighter absolute magnitudes than the value $M_{\rm
crit}(r)$ from equation (\ref{eq:defsMcrit}) will enter the catalogue. After
a large number $N$ of values of $\vtheta$ have been explored, we have
 \[
\phi(\vJ)\simeq{1\over N}\sum_{i=0}^N\int_{-\infty}^{M_{{\rm crit}}(r_i)}\d
M\,F(M).
\]
 Since $\phi$ does not depend on $f$, it needs to be evaluated only at the
start of the procedure for optimising $f$.  

The survey contains no information about the value of $f(\vJ)$ at actions for
which $\phi(\vJ)=0$, so it is immaterial whether we change the value of $f$
at these actions. When we adopt a functional form for $f$, we are in fact
varying $f$ in these invisible regions, and thus at the end of the process
arrive at a prediction for that can be tested by a deeper or wider survey.

\subsection{Binning the data}\label{sec:binning}

The computational cost of evaluating the likelihood of a catalogue scales as
the product of the number of stars in the catalogue and the number of tori
used to sample the model. Even for a catalogue that contains $\sim10^5$
stars, the cost is high. We now investigate the scope for reducing the cost
by binning the data.

The log likelihood (\ref{eq:trial}) is a sum of the contributions from
individual stars. Since the contribution from an
individual star is practically unchanged by shifting its point $\vu$ in data
space by a small amount within the core of its error ellipsoid, the
calculated likelihood of the catalogue will not change much if we group stars
with data points that lie within a small cell in data space and attribute to
all of them the contribution to $L$ from the centre of the bin. A slightly
refined version of this basic idea is as follows.

We establish a Cartesian grid in data space, the cell spacing in most
dimensions being of order half the typical observational uncertainty of the
corresponding observable. If we applied this criterion to $l$ and $b$, we
would obtain an absurdly fine grid on the sky. Therefore, in $l$ and $b$ we
take the separation between bins to be comparable to the changes in angle
over which we expect the distribution of stars to change significantly --
this might be as large as $10^\circ$ in $l$ and a degree or so in $b$ at low
latitudes and larger increments near the poles.  Having established our grid,
we assign stars to cells. Finally we evaluate $\pr$ at the centre of each
cell and form $L$ by adding the logarithm of this value times the number of
stars assigned to that cell. The saving in computation from binning is clearly
proportional to the number of stars assigned to each cell, which increases
with the adopted bin sizes and decreases with the number of quantities
actually observed. Hence binning is most attractive for the sparsest data
set, namely measurements of $(b,l,m,\vmu)$. However, this is already a
five-dimensional space. For a survey of a third of the sky, it might be
possible to get by with 100 bins on the sky. A few tens of bins would be
required for the apparent magnitudes, and for each component of $\vmu$ ten
bins might suffice. Thus a minimal grid will have $>10^5$ bins, so even
with the simplest conceivable data, binning first becomes advantageous when the
number of stars in the catalogue exceeds a million. In the case of Gaia, the
list of observables must be expanded to include at least $\varpi$, and the
number of bins required to do justice to the proper-motion data must be
increased by a factor of at least 10, implying a grid with $>10^6$ bins.  In
reality one would want to include colour data, and some line-of-sight
velocities, and the number of bins would be pushed up to $\sim10^9$,
essentially the number of stars in the catalogue. 

Thus on account of the high dimensionality of the problem posed by current
and future surveys of the Galaxy, the prospects for binning reducing the
computational cost of fitting models are not bright. 

\section{Optimising the DF}\label{sec:optimise}

The formulae of the last section are used to evaluate the log likelihood $L$
of a catalogue given a particular model. In this section we explain how we
estimate the probability distributions of the parameters that appear in the
\df. We do this with the Markov Chain Monte-Carlo (MCMC) algorithm: we choose some
plausible set of parameters $\va$ for the \df\ and evaluate $L$ for these
parameters. Then we generate a random change $\delta\va$ in the parameters
and evaluate $L$ for the \df\ with parameters $\va'\equiv\va+\delta\va$. If
the new value $L'$ exceeds the old value $L$, we set $\va=\va'$, while if
$L'<L$ we set $\va=\va'$ with probability $\exp(L'-L)$. After a sufficient
time, the resulting sequence of values of $\va$ sample the underlying \pdf\
of $\va$ \citep[e.g.\ \S4.3 of][]{TCP}.

Thus our approach to model fitting employs Monte-Carlo sampling in two
distinct ways. To evaluate $L$ for any given parameters $\va$ we evaluate the
action-space integral by Monte-Carlo sampling action space, and to determine
the \pdf\ of the parameters of $f$ we Monte-Carlo sample parameter space.

It is useful to keep a record of the individual contributions to the sums
over $k$ in equation (\ref{eq:short}) for the following reason.  Once the
likelihood of the data given the current \df\ $f$ has been evaluated, a
neighbouring \df, $\widetilde f$, is chosen, and the likelihood is
re-evaluated. The new value of any sum over $k$ can be obtained by,
in equation (\ref{eq:short}), making the replacement
 \begin{equation}\label{eq:replacedf}
\sum_k \longrightarrow \sum_k[\widetilde f(\vJ_k)/f(\vJ_k)].
\end{equation}
Consequently, if when we first evaluate the likelihood of
the catalogue we record for each star the contribution that each torus makes
to the probability of seeing that star, we can evaluate the likelihood of the
catalogue for any other \df\ at speed. The price of this speed is having to
record an array of size the number of stars in the catalogue times the number
of tori employed. Fortunately, the array is somewhat sparse because a
significant fraction of tori make a negligible contribution to the likelihood of a
given star. The more precise the data are, the sparser the array becomes.

A key to computational efficiency on the first evaluation of the likelihood
of the catalogue is to identify in advance for each star the limited number
of tori that contribute significantly to the integral in equation
(\ref{eq:short}). To do so efficiently we create, for each torus, a grid in
$l,b$ and heliocentric $r$, and store the ranges of possible $v_l$, $v_b$ and
$v_{\rm los}$ for each bin in the grid that the torus goes through (found to
a good approximation by sampling at many $\vtheta$ from the torus). With this
information we can then, for each star, quickly discard tori which are
clearly not relevant because the bins associated with the appropriate line of
sight are either empty or have $r$ and/or velocity ranges which are clearly
incompatible with the observations.

For a catalogue of significant size the extra computational effort required
to produce these grids is small compared to the computational effort saved by
reducing the number of integrals that need to be performed because a grid
only needs to be computed once per torus, rather than once per star per
torus.  In the examples discussed in Section \ref{sec:test} the grids
reduce the number of line-of-sight integrals which need to be found by
between a factor of $\sim$$10$ when only $\vmu$ is provided, and a factor of
$\sim$$50$ when $\vmu$, $v_{\rm los}$ and $\varpi$ are provided.

\subsection{How many tori do we need?}

As just explained, estimates of $L$ for many different parameter values are
obtained with the same set of representative tori, and the final optimum
values of the parameter and their uncertainties are usually based on a single
set of sampling tori. Unless this set is sufficiently rich, the parameter
estimates are likely to be biased in the sense that their optimum values
differ from the true values by more than the returned uncertainties: the
latter reflect both the input observational errors and the statistical
uncertainty arising from the finite number of stars in the catalogue we are
fitting. They do not include statistical error associated with the use of a
finite number of tori to evaluate integrals over action space. 

So, how many
tori do we need to use? It is evident that at least one torus must yield a
point $\vu'$ in the space of observables that lies within $\sim1\sigma$ of
the measured location of each star. Stars
for which only one torus contributes points to the $1\sigma$ error ellipsoid
tend to bias the parameter values: in order to make such stars probable the
computer has to make that torus probable; if this torus differs significantly
from the star's actual torus, the probability associated with that star is
being misplaced within action space, and we infer an incorrect \df. For
safety we require several tori to contribute points to the $1\sigma$ error
ellipsoid, for then the greatest weight can be assigned to whichever one of
them lies close to the star's true torus. Clearly the higher the quality of
the catalogue, the smaller are the error ellipsoids and the more tori are
needed to fulfil the criterion that several tori contribute to every
$1\sigma$ error ellipsoid.

We can quantify this idea as follows. Equation (\ref{eq:short})
gives the probability of finding a star with observables $\vu$ as a sum of
contributions from individual tori. If we define
 \begin{equation}\label{eq:defspk}
p_k\equiv {\int\!\d r'\left|{\p(\vtheta)\over\p(b,l,r')}\right|
\int\d
M\,F(M)G_3^7(\vu,\vu'_k,\vsigma)\over\pr(\vu|f)\sum_{k'}\phi(\vJ_{k'})},
\end{equation}
 then we have $\sum_k^N p_k=1$ and the contribution of the $k$th torus to
$\pr(\vu|f)$ is proportional to $p_k$. The Shannon entropy of the probability
distribution $\{p_k\}$,
 \begin{equation}
S(\vu|f)=-\sum_k^N p_k\ln p_k,
\end{equation}
 is the natural measure of the extent to which the probability $\pr(\vu|f)$
is contributed by a large number of terms in the sum or dominated by a single
largest contribution: $S$ vanishes if there is one dominant contribution, so
$p_k=\delta_{kk_0}$, and peaks at $S=\ln N$ when $p_k=1/N$ for all $k$;
if there are just $n\ll N$ tori that might plausibly produce the data for a given
star, $n$ of the  $p_k$ will be of order $1/n$ and the rest will vanish, so
the entropy will be of order $S\simeq\ln n$. Hence $\e^S$ is quite generally
an estimate of the number of tori that are consistent with the star's data.
Using too few tori to conduct the sum in equation (\ref{eq:short}) is
signalled by some stars having values of $\e^S$ smaller than unity/a few and
results in the formal errors on the parameters being too small, with the
result that the true parameters lie outside the $\sim1.5\sigma$ error
ellipsoid. 

While having $\e^S\sim1$ suggests that too few tori are being used, the
following argument shows that a larger value of $\e^S$ does not guarantee
that enough tori are being used to give reliable results. Two small samples
of tori drawn from different \df s can by chance have almost identical
distributions in action space. Suppose this distribution happens to fit the
data perfectly. Then the two different parent \df s will appear to maximise
the likelihood of the data, depending on which \df\ the tori were in fact drawn
from. The best way to check that enough tori are being used is to draw a new
sample of tori from the \df\ that maximises the likelihood with the original
sample of tori, and then to repeat the maximisation process using the new
tori in the analysis. If enough tori were used, the new pdf of the \df\ will
differ negligibly from the old one.

\section{Tests}\label{sec:test}

In this section we test the ability of our procedure to recover from a mock
catalogue the \df\ from which the catalogue was obtained. Hence we (i)
build a dynamical model with known \df, (ii) draw a sample of
pseudodata from this model by ``observing'' it from the location of the Sun
and then add errors to the ``observations'', and (iii) use the algorithm
described in Section~\ref{sec:optimise} to constrain the \df. In this final
step we assume the correct functional form for the \df\ and enquire how
accurately we can recover from the pseudodata the values of the \df's
parameters. We take the distance from the Galactic Centre to the Sun to be
$R_0=8.5\kpc$.

\subsection{The adopted luminosity function} \label{sec:LF}

\begin{figure}
  \centerline{\resizebox{\hsize}{!}{\includegraphics[angle=270]{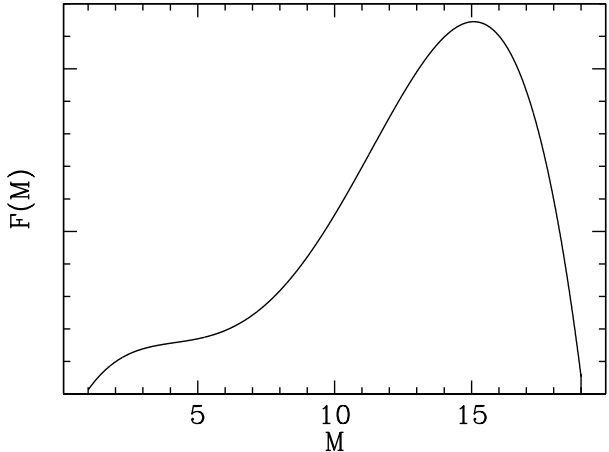}}}
  \caption{Luminosity function described in Section~\ref{sec:LF}, and used in 
    all tests.}
\label{fig:LF}
\end{figure}

Since the velocity we infer for a star from its proper motion is proportional
to the star's distance, whatever distance information we have is going to be
crucial for the model-fitting process. For simplicity, we
are assuming that the Galaxy contains a single stellar population, and we are
not using colour information. Consequently, it suffices to specify the
luminosity function. We use a simple polynomial approximation to the general 
$V$-band luminosity function described in
\cite{GalacticAstronomy} Table 3.16:
 \begin{equation} \label{eq:LF}
 F(M) \propto \left\{
\begin{array}{ll} 
-14.9 + 21\,M -5.4\,M^2 \\
\qquad+ 0.59\,M^3 -0.019\,M^4 & \hbox{ for } 1<M<19 \\
0 & \mathrm{otherwise}. \\ 
\end{array}
\right.
\end{equation}
This function is plotted in Figure~\ref{fig:LF}. 

\subsection{The adopted gravitational potential}

We use the Galactic potential which \cite{McMillan2011} gives for a
``convenient'' Galaxy model.  This axisymmetric model consists of a Galactic bulge,
thin and thick exponential discs, and a \cite*{NFW1996} halo. The potential
defines the Local Standard of Rest (LSR) and we assume that the Sun's
velocity with respect to the LSR is that given by \cite{SchoenrichBD}.

\subsection{The adopted distribution function}

We have run tests using two model Galaxies, both based on a 
``quasi-isothermal'' \df\ \citep{BinneyM11}
 \begin{equation}\label{totalDF}
f(J_r,L_z,J_z)=f_{\sigma_r}(J_r,L_z)\times{\nu_z\over2\pi\sigma_z^2}
\,\e^{-\nu_z J_z/\sigma_z^2},
\end{equation}
where
 \begin{equation}\label{planeDF}
f_{\sigma_r}(J_r,L_z)\equiv{\Omega\Sigma\over\pi\sigma_r^2\kappa}\bigg|_{\Rc}
[1+\tanh(L_z/L_0)]\e^{-\kappa J_r/\sigma_r^2}.
\end{equation}
 Here $\Omega(L_z)$ is the circular frequency for angular momentum $L_z$,
$\kappa(L_z)$ is the radial epicycle frequency and $\nu(L_z)$ is its vertical
counterpart.  $\Sigma(L_z)=\Sigma_0\e^{-(R-\Rc)/R_\d}$ is the (approximate)
radial surface-density profile and we set $R_\d=3\kpc$, 
where $\Rc(L_z)$ is the radius of the
circular orbit with angular momentum $L_z$. The factor $1+\tanh(L_z/L_0)$ in
equation (\ref{planeDF}) is there to effectively eliminate stars on
counter-rotating orbits; the value of $L_0$ is unimportant in this study 
provided it is
small compared to the angular momentum of the Sun. In equations
(\ref{totalDF}) and (\ref{planeDF}) the functions $\sigma_z(L_z)$ and
$\sigma_r(L_z)$ control the vertical and radial velocity dispersions. The
observed insensitivity to radius of the scale-heights of extragalactic discs
motivates the choices
 \begin{eqnarray}\label{eq:sigmas}
\sigma_r(L_z)&=&\sigma_{r0}\,\e^{q(R_0-\Rc)/R_\d}\nonumber\\
\sigma_z(L_z)&=&\sigma_{z0}\,\e^{q(R_0-\Rc)/R_\d},
\end{eqnarray}
 where $q=0.45$ and $\sigma_{r0}$ and $\sigma_{z0}$ are approximately equal
to the radial and vertical velocity dispersions at the Sun. For discussion of
the value of $q$ and these choices of functional form, see B10.

The first distribution function we consider is a single thin 
quasi-isothermal disc, which  was constructed with $R_\d=3\kpc$, 
$L_0=10\kpc\kms$ and $\sigma_{r0}=\sigma_{z0}=10\kms$. This is  
rather dynamically cold, but we use it as a simple example to
demonstrate and test the basic principles of the apparatus.

We have also tested the apparatus on a more realistic model Galaxy that has
both thin and thick discs, with the thick disc contributing 23 per cent of the
total disc surface density at the Sun.  Table \ref{tab:DF} gives the values
of the parameters in the \df\ of this model.

B10 showed that by superposing a large number of \df s of a 
very similar form to these quasi-isothermal \df s, one can obtain a 
model that is consistent with the local stellar density and velocity 
distribution. In 
this study we, like \cite{BinneyM11}, restrict ourselves to these simple 
one- or two-disc models in order to provide some straightforward 
demonstrations of the principles discussed in Section~\ref{sec:strategy}. 
Extending this work to the more complicated \df s used by B10
is in principle straightforward.

\begin{table}
\caption{The parameters of the \df\ of the model that contains both thin and
thick discs}\label{tab:DF}
\begin{tabular}{lcccc}
Disc&$R_\d/$kpc&$\sigma_{r0}$/km$\,{\rm s}^{-1}$
&$\sigma_{z0}$/km$\,{\rm s}^{-1}$&$L_0$/kpc\,km$\,{\rm s}^{-1}$\\
\hline
Thin & 3.0 & 27 & 20 & 10\\
Thick & 3.5 & 48 & 44& 10\\
\end{tabular}
\end{table}

\subsection{Constructing a pseudo-catalogue}\label{sec:pseudodata}

\begin{figure*}
\centerline{
\includegraphics[width=.3\hsize]{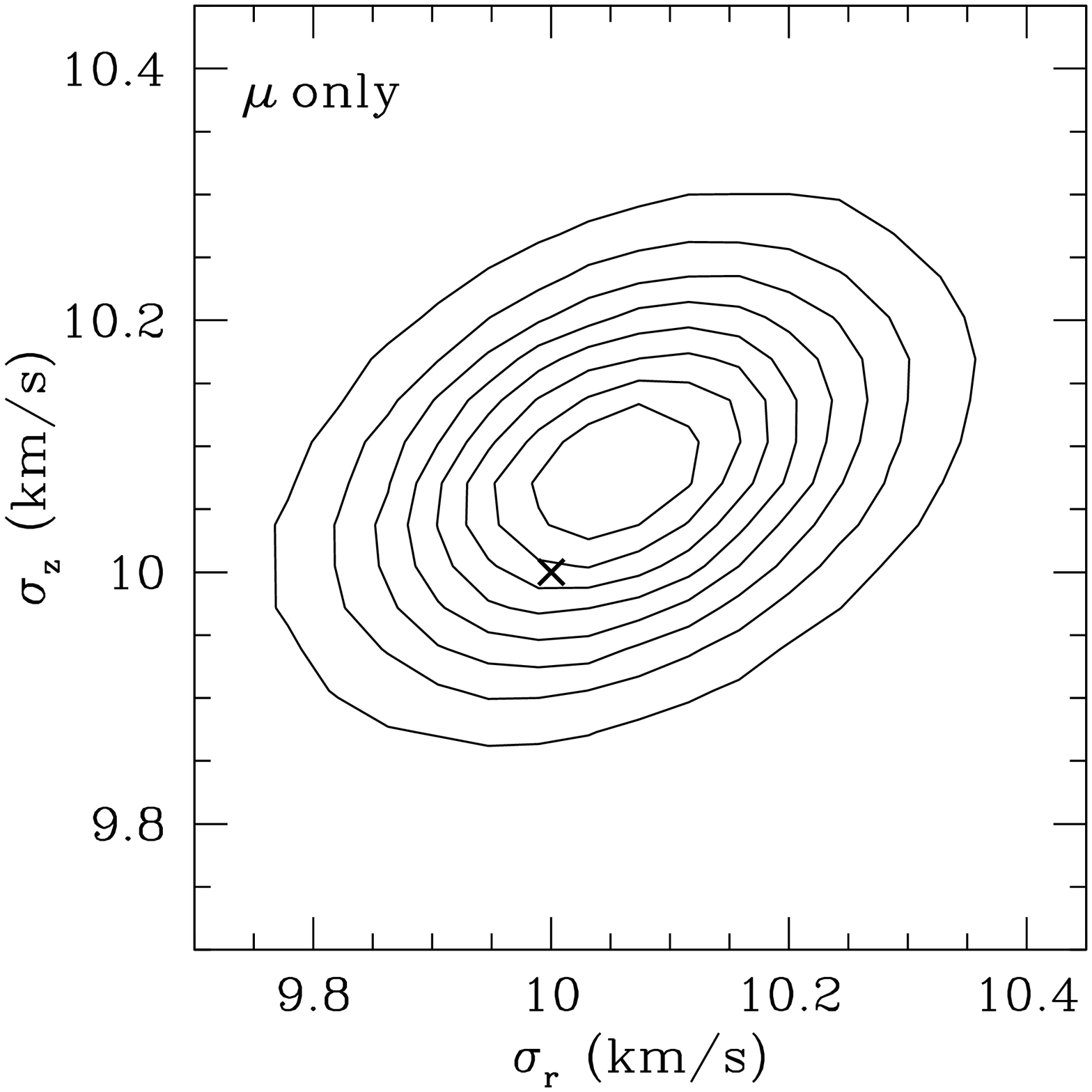}
\includegraphics[width=.3\hsize]{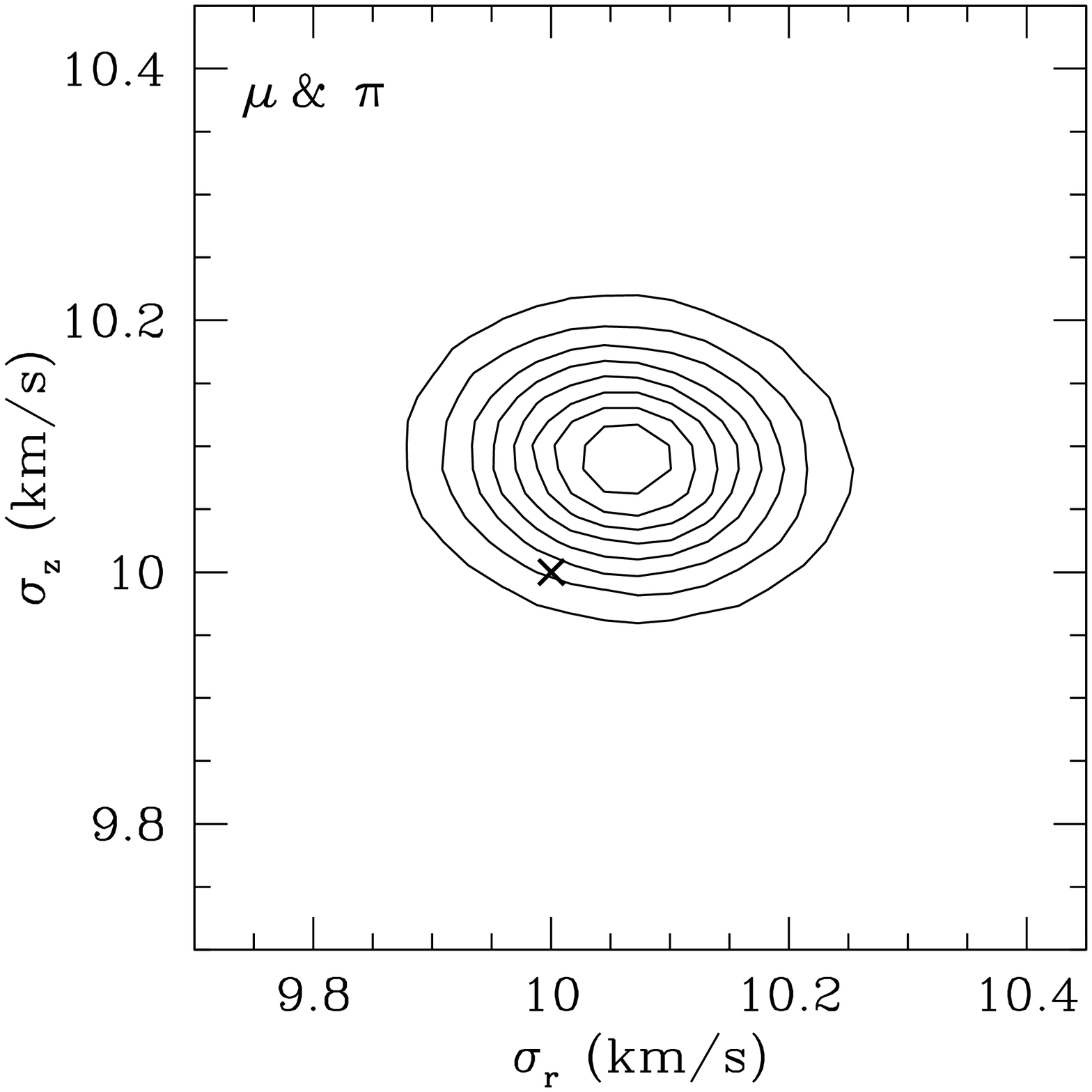}
\includegraphics[width=.3\hsize]{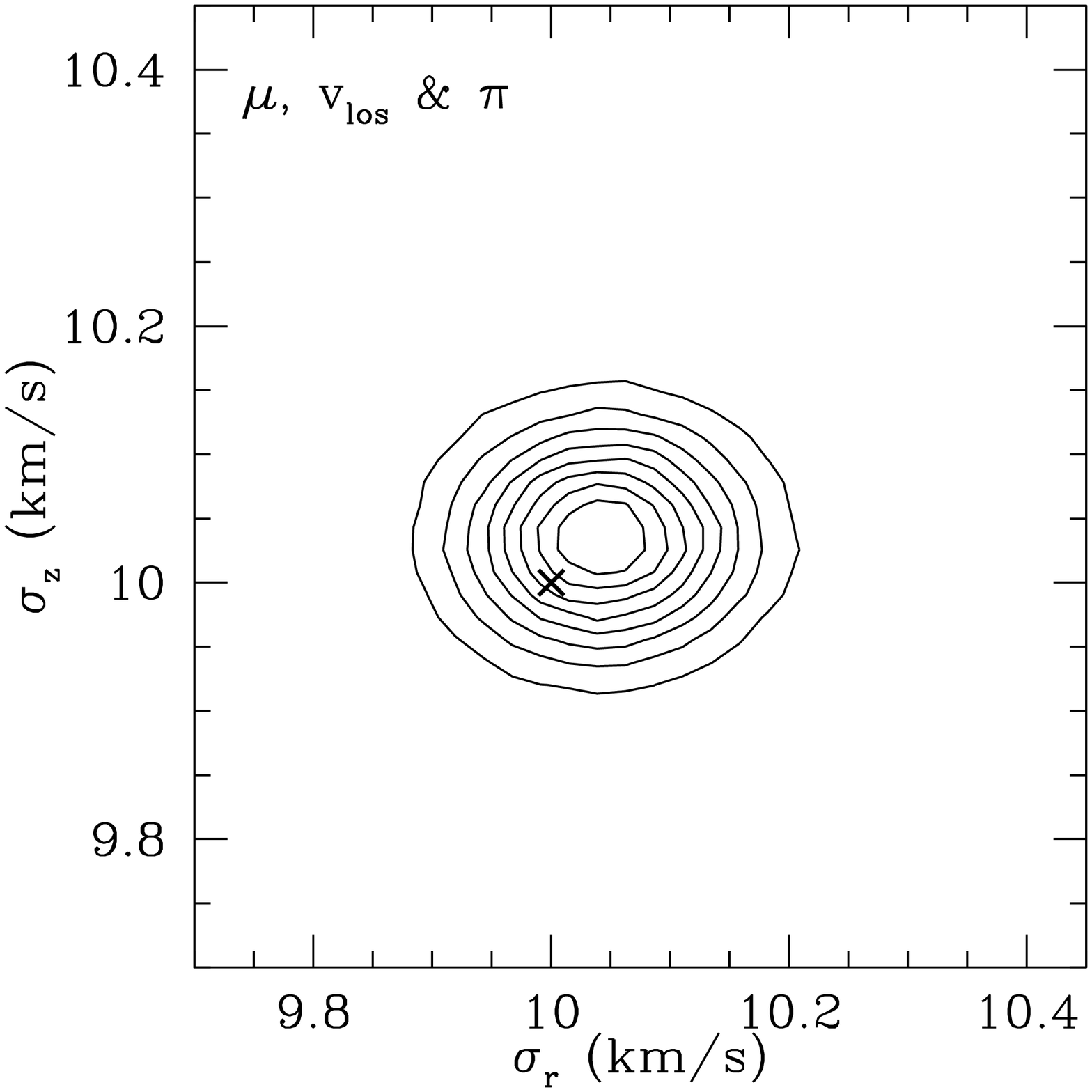}
}
 \caption{Tests of our ability to recover the parameters of the \df\ a
catalogue of $5\,000$ stars. Each panel shows the pdf of the \df\ in the
$(\sigma_{r0},\sigma_{z0})$ plane. The data from which these pdfs were
recovered included photometry and proper motions in all three cases. The
extreme left panel gives the pdf when no additional data are available, the
centre panel shows what is achieved by adding parallaxes, and the rightmost
panel shows the pdf when both parallaxes and line-of-sight velocities are
available. Contour levels are at $n/9$ of the peak probability density, where
$n=1,\ldots,8$. The cross marks the true value of the
parameters.} \label{fig:onedisc}
\end{figure*}

\begin{table*}
\caption{Means and dispersions in $\!\kms$ of parameters of the \df\
recovered from data made with just a thin disc. The first three columns give
results for a catalogue with $5\,000$ stars in a catalogue with Gaia-like
measurement errors. The last two columns give
the uncertainties arising from finite catalogue size with $5\,000$ and $10\,000$
stars in an error-free catalogue.}\label{tab:onedisc}
\begin{tabular}{lcc|ccc}
&\multispan3{\hfil Gaia-like errors\hfil}&\multispan2{\hfil perfect data\hfil}\\
&$\mu$ only&$\mu,\varpi$&$\mu,\varpi,v_{\rm los}$&$5\,000$ stars&$10\,000$ stars\\
\hline
$\sigma_{r0}$ &$10.06\pm0.14$&$10.06\pm0.09$&$10.044\pm0.075$&$10.02\pm0.07$&$10.06\pm0.05$\\
$\sigma_{z0}$ &$10.08\pm0.10$&$10.09\pm0.06$&$10.03\pm0.06$&$9.97\pm0.05$&$9.97\pm0.04$\\
\end{tabular}
\end{table*}

To construct a pseudo-catalogue of stars, we first randomly sample the
distribution function in action space.  If we take care to choose $f$ so it
can be evaluated without actually constructing a torus (which requires that
it does not contain exact orbital frequencies) very little work is involved
in evaluating $f(\vJ)$ so MCMC sampling, in which at best a third of the
points investigated are actually selected, is not expensive. For each chosen
$\vJ$ a torus is constructed.

Our tests are performed with a pseudo-catalogue that is magnitude limited
with maximum apparent magnitude $m=17$, and we confine the pseudo-catalogue
to $b>30^\circ$ because we have neglected extinction.  The shape of the
survey volume is then comparable to that of recent surveys (e.g. SDSS, RAVE).
For each computed torus we evaluate the selection function $\phi(\vJ)$ as
described in Section \ref{sec:basic}.  For some of our selected values of
$\vJ$, the selection function will vanish because the torus does not
intersect the survey volume. Such tori may be excluded from the list and it
is in fact convenient to identify many of these tori in advance of computing
them.  Specifically, for a reasonable number of points in the $J_r , J_\phi$
plane we find the minimum value of $J_z$ that takes a torus with given $J_r$
and $J_\phi$ into the survey volume -- this is conveniently done using the
adiabatic approximation \citep{BinneyM11,SchoenrichB11}. Then by interpolating between these
values we can assess whether a given point $\vJ$ is well outside the region
of action space that can contribute to the catalogue.  

To add a star to the catalogue, we select a torus from our list at random.
Then we choose a value for each $\theta_i$ at random between $0$ and $2\pi$,
and an absolute magnitude $M$ randomly from our luminosity function
(eq.~\ref{eq:LF}). This defines the position, velocity, and apparent
magnitude of a star. Observational errors are then added to the observational
data for each star. The assumed errors were $0.2\mas$ in parallax,
$0.2\mas\yr^{-1}$ in proper motion, and $5\kms$ in line-of-sight velocity.
The star enters the catalogue if its observational data place it within
our survey limits, and is discarded otherwise. We repeat this process until
we have produced a catalogue of the desired size.

\subsection{Case of only a thin disc}\label{sec:onedisc}

Fig.~\ref{fig:onedisc} shows the pdfs of the parameters $\sigma_{r0}$ and
$\sigma_{z0}$ for the pure thin-disc model when there are data for $5\,000$
stars and only photometry and proper motions are available (leftmost panel),
or parallaxes are also available (central panel), or both parallaxes and
line-of-sight velocities are available (rightmost panel). The means and
dispersions of these pdfs are given by the last three columns of
Table~\ref{tab:onedisc}. We see that in all three cases, the true parameter
values lie within the expected error ellipse and the uncertainties of the
parameters are $<1.5$ per cent. We obtained these results using $40\,000$ tori
with non-vanishing values of $\phi(\vJ)$.

The situation when neither parallaxes nor line-of-sight velocities are
available is of particular interest.  Given the breadth of the luminosity
function plotted in Fig.~\ref{fig:LF}, it is remarkable that the parameters
can be determined with such precision because to do this, the algorithm has
to correctly infer the typical velocities of stars, and it can deduce a
velocity from a given proper motions only by correctly inferring the
distance. By itself the broad luminosity function does not provide sufficient
distance information. The required distance information is inferred by
piecing together several lines of evidence. For example, in the given
potential, a model with larger values of $\sigma_{r0}$ and $\sigma_{z0}$
would have a thicker disc, so stars seen at $b=30^\circ$ along $l=0$ and
$l=180^\circ$ would typically have Galactocentric radii that differed more
than in a model with small $\sigma_{i0}$. Consequently, the ratio of the
stellar densities seen along $l=0$ and $l=180^\circ$ increases with the
$\sigma_{i0}$, and from the ratio present in the data the algorithm can
constrain the $\sigma_{i0}$.  An independent constraint is provided by the
Sun's motion with respect to the LSR, since both the asymmetric drift of the
disc and the typical distance to stars vary with the $\sigma_{i0}$, so the
dispersions are constrained by the mean proper motion of disc stars. This is
essentially the classical concept of secular parallax \citep[e.g. \S2.2.3
of][]{GalacticAstronomy} in a sophisticated context.

When only proper motions are available, the error ellipse is tilted with
respect to the axes in the sense of a positive correlation between
$\sigma_{r0}$ and $\sigma_{z0}$.  The correlation arises because increasing
$\sigma_{z0}$ without increasing $\sigma_{r0}$ would, for example, at
$l\simeq90^\circ$ increase $\mu_b$ without increasing $\mu_l$,
and this would be apparent in the data. The impact on the data of increasing
both dispersions in step can be to some extent masked by increasing the
assumed distances to stars. The central panel of \figref{fig:onedisc} shows
that adding parallaxes eliminates this correlation. The panel on the extreme
right of \figref{fig:onedisc} shows that for this model little is gained by
adding line-of-sight velocities with errors of $5\kms$, presumably because
the random motions of the stars are so small.

\begin{figure}
\centerline{\includegraphics[width=.7\hsize]{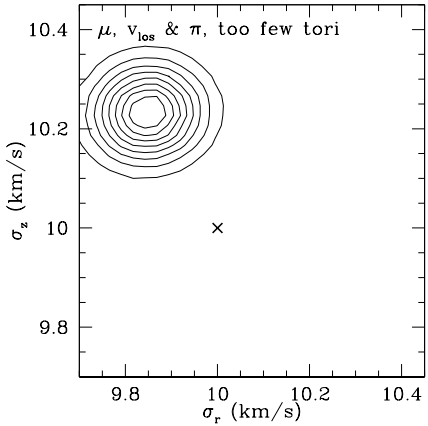}}
\caption{A bias introduced by using too few tori. Here only 2\,000 tori have
been used to recover the parameters of the \df\ of the single-disc model.}
\label{fig:fewtori}
\end{figure}

\subsubsection{Case of too few tori}

In Section \ref{sec:optimise} we noted that using too few tori is liable to
produce biased results. This phenomenon is illustrated by
\figref{fig:fewtori}, which shows the same results as the rightmost panel of
\figref{fig:onedisc} except that 2\,000 rather than $40\,000$ tori have been
used in the analysis.  The algorithm deduces that
$\sigma_{r0}=(9.85\pm0.07)\kms$, $\sigma_{z0}=(10.23\pm0.06)\kms$ so the
central values lie over $3\sigma$ from the truth.

\subsubsection{Choice of trial distribution function}

\begin{figure}
\centerline{\includegraphics[width=.7\hsize]{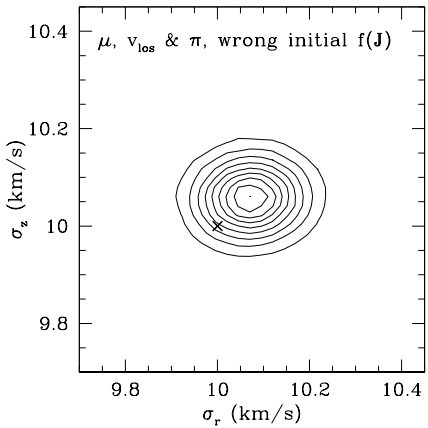}}
\caption{The pdf in the $(\sigma_{r0},\sigma_{z0})$ plane for the same  
catalogue as that used in rightmost panel of Fig.~\ref{fig:onedisc}, when
an incorrect trial \df\ (with $\sigma_{r0}=\sigma_{z0}=12\kms$) was 
used to perform the original integral over $\vJ$.}
\label{fig:wrongdf}
\end{figure}

\begin{figure}
\centerline{\includegraphics[width=.7\hsize]{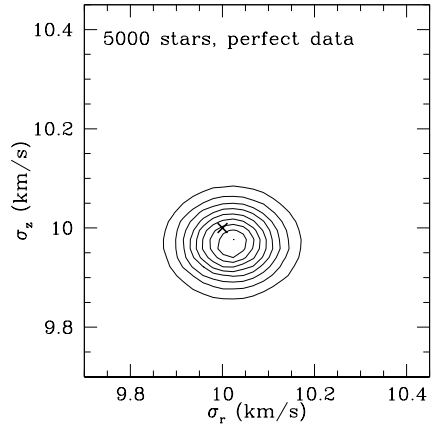}}
\centerline{\includegraphics[width=.7\hsize]{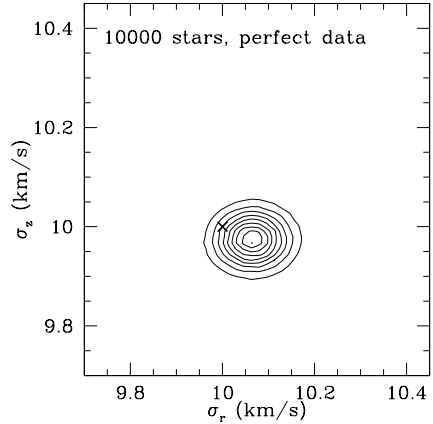}}
\caption{The pdf in the $(\sigma_{r0},\sigma_{z0})$ plane when the only
source of uncertainty is the finite size of the catalogue.}
\label{fig:noerror}
\end{figure}

The Monte-Carlo integrals over $\vJ$ in equation (\ref{eq:finalP}) are 
performed using an initial trial \df, which then allows us to evaluate 
the likelihood
of similar \df s using the substitution given in
equation~(\ref{eq:replacedf}). 
In the other cases reported in this study we use the \df\ from which the stars 
were drawn as this trial \df, for convenience. To use this method in practice
we need to be confident that it returns a correct result when a 
different trial \df\ is used. 

In Fig.~\ref{fig:wrongdf} we show the pdf in 
the $(\sigma_{r0},\sigma_{z0})$ plane found for the thin-disc \df\ using the 
same input catalogue as that used to produce the rightmost panel of 
Fig.~\ref{fig:onedisc} and the rightmost column of 
Table~\ref{tab:onedisc}, but with an initial trial \df\ which is a 
``quasi-isothermal'' disc with $\sigma_{r0}=\sigma_{z0}=12\kms$. The 
algorithm finds the results $\sigma_{r0}=(10.07\pm0.08)\kms$, 
$\sigma_{z0}=(10.06\pm0.06)\kms$, which is very close to the result found with 
the correct trial \df. Using a trial \df\ which differs from the true \df\ 
does reduce the effective number of tori which are used to determine the 
true \df, as it places an excessive number of tori in some regions of 
limited interest in action space, leaving a reduced number of tori in 
the relevant regions in action space. If the \df\ determined by the algorithm 
differs significantly from the trial \df\ it is sensible to use the \df\ that
is determined as a new initial trial \df, and repeat the analysis to ensure 
that the results are not biased by the low effective number of tori.

\subsubsection{Irreducible statistical error}

\begin{figure*}
\centerline{
\includegraphics[width=.3\hsize]{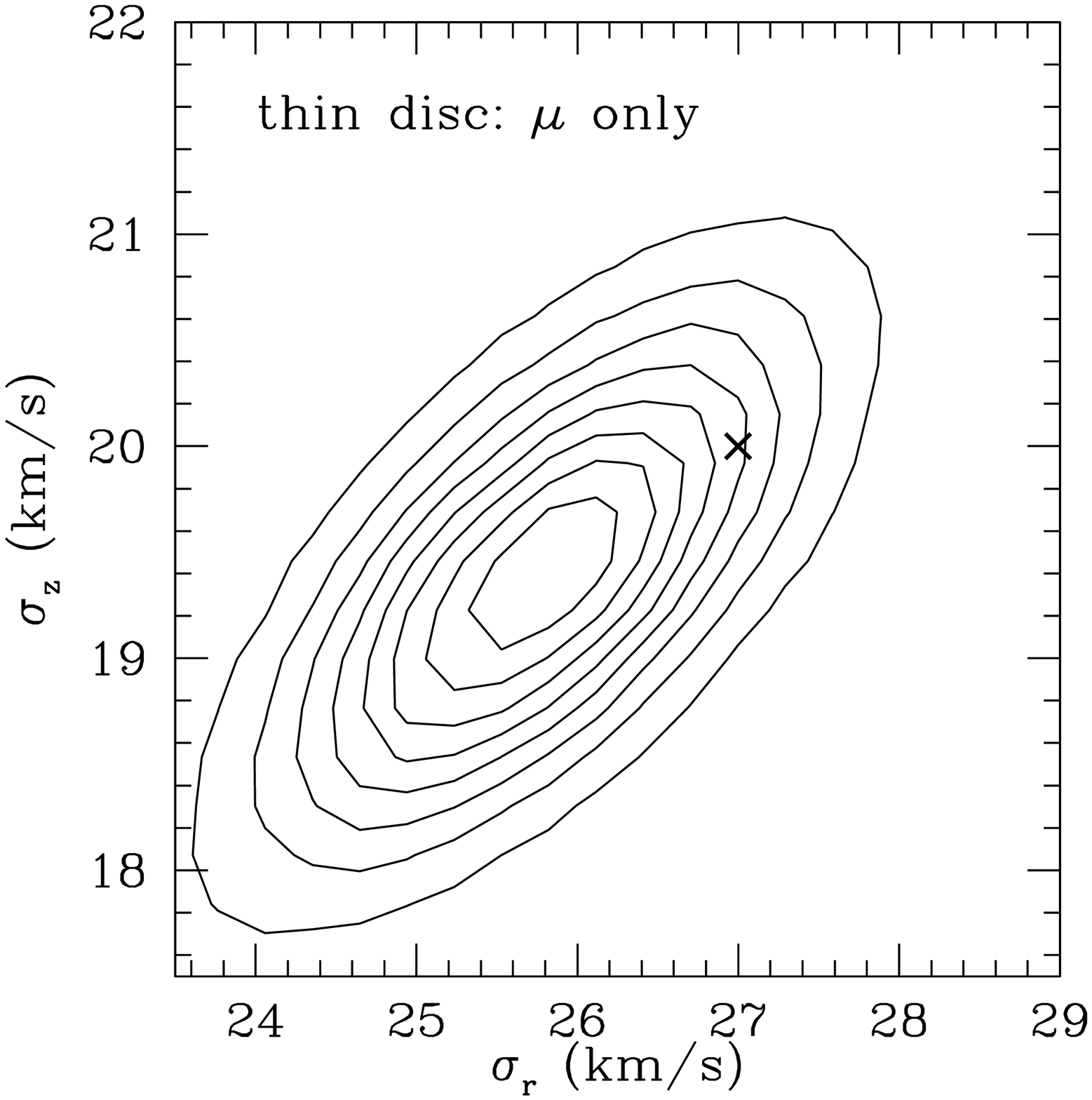}
\includegraphics[width=.3\hsize]{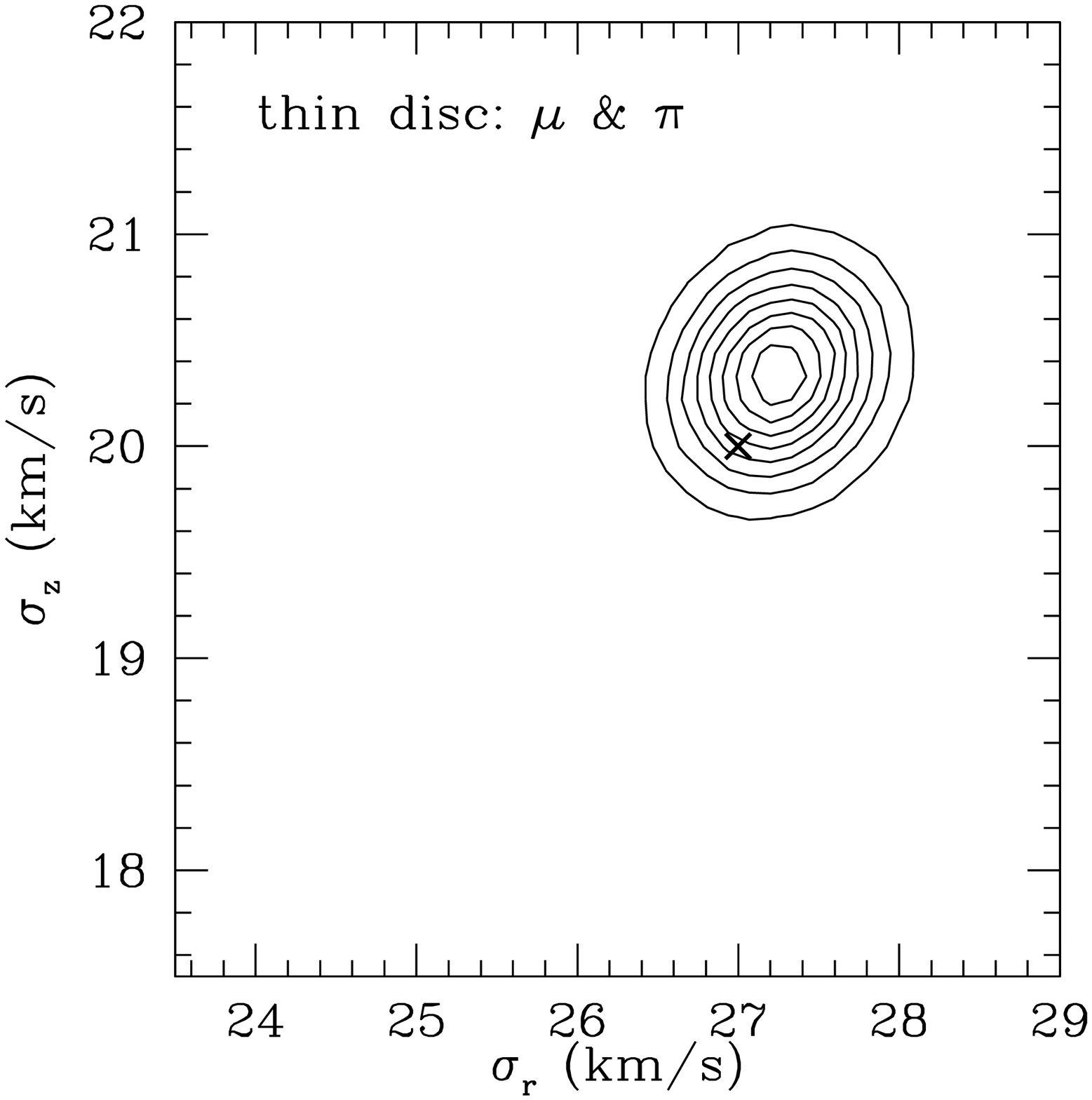}
\includegraphics[width=.3\hsize]{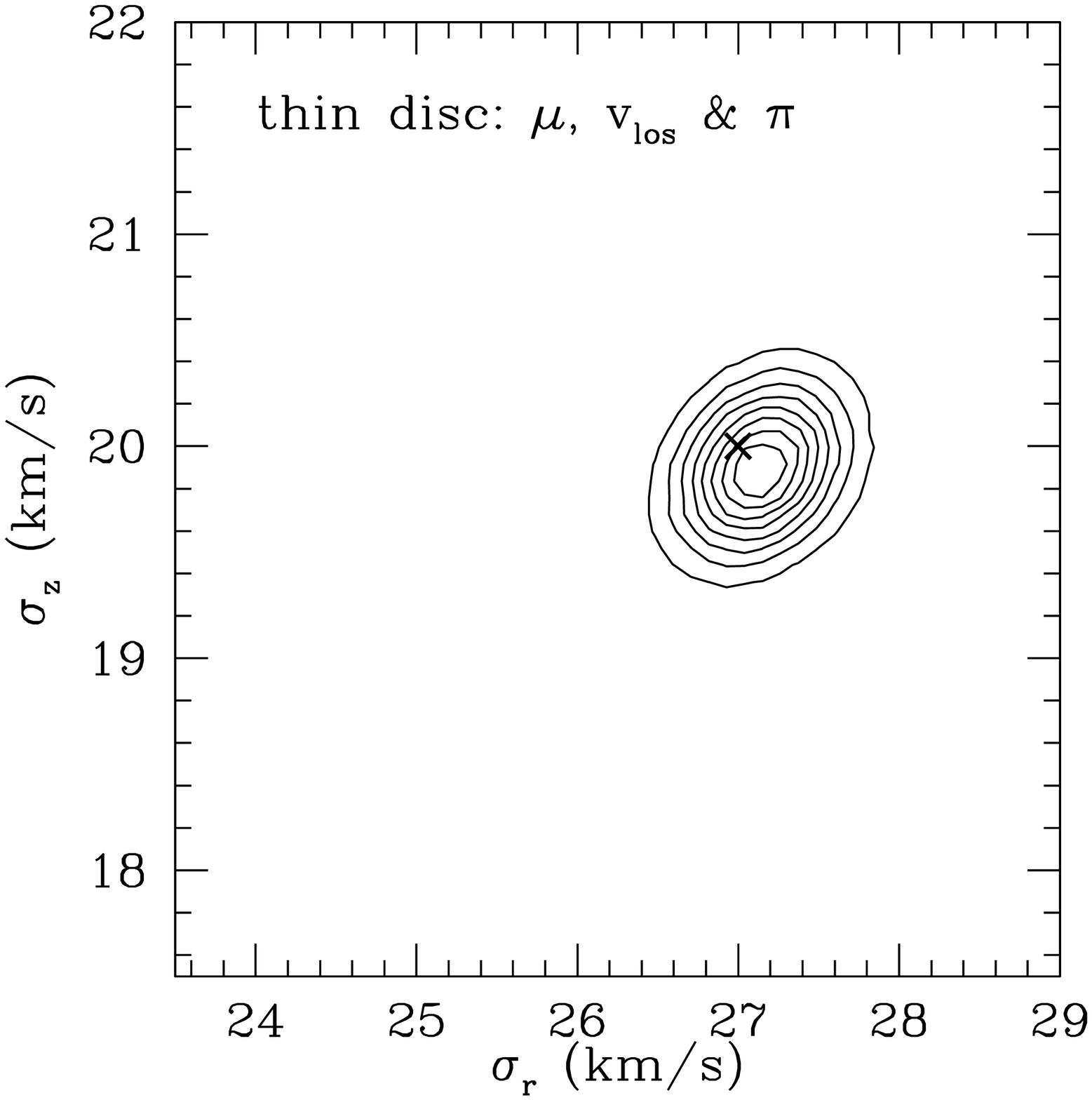}
}
\centerline{
\includegraphics[width=.3\hsize]{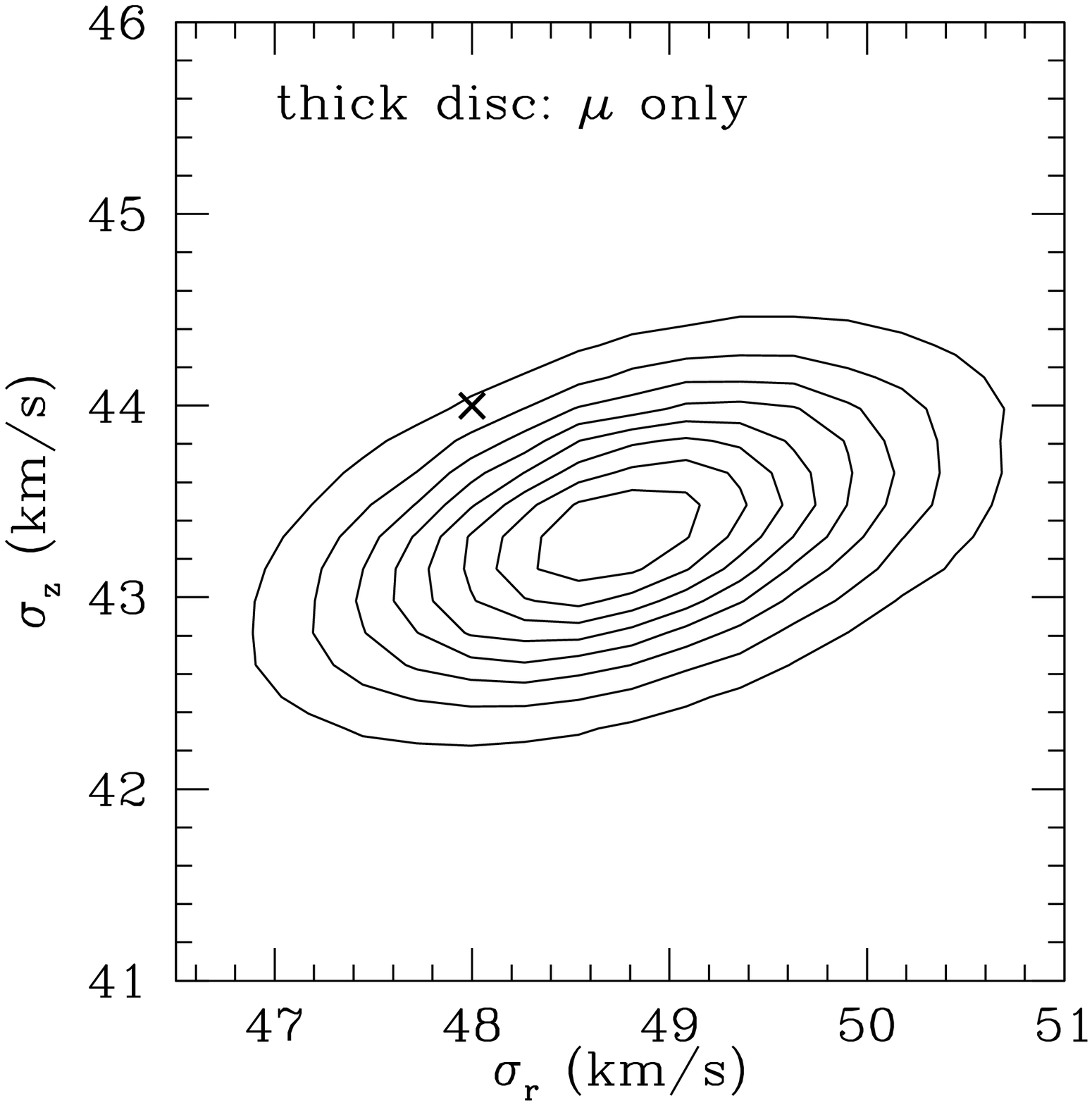}
\includegraphics[width=.3\hsize]{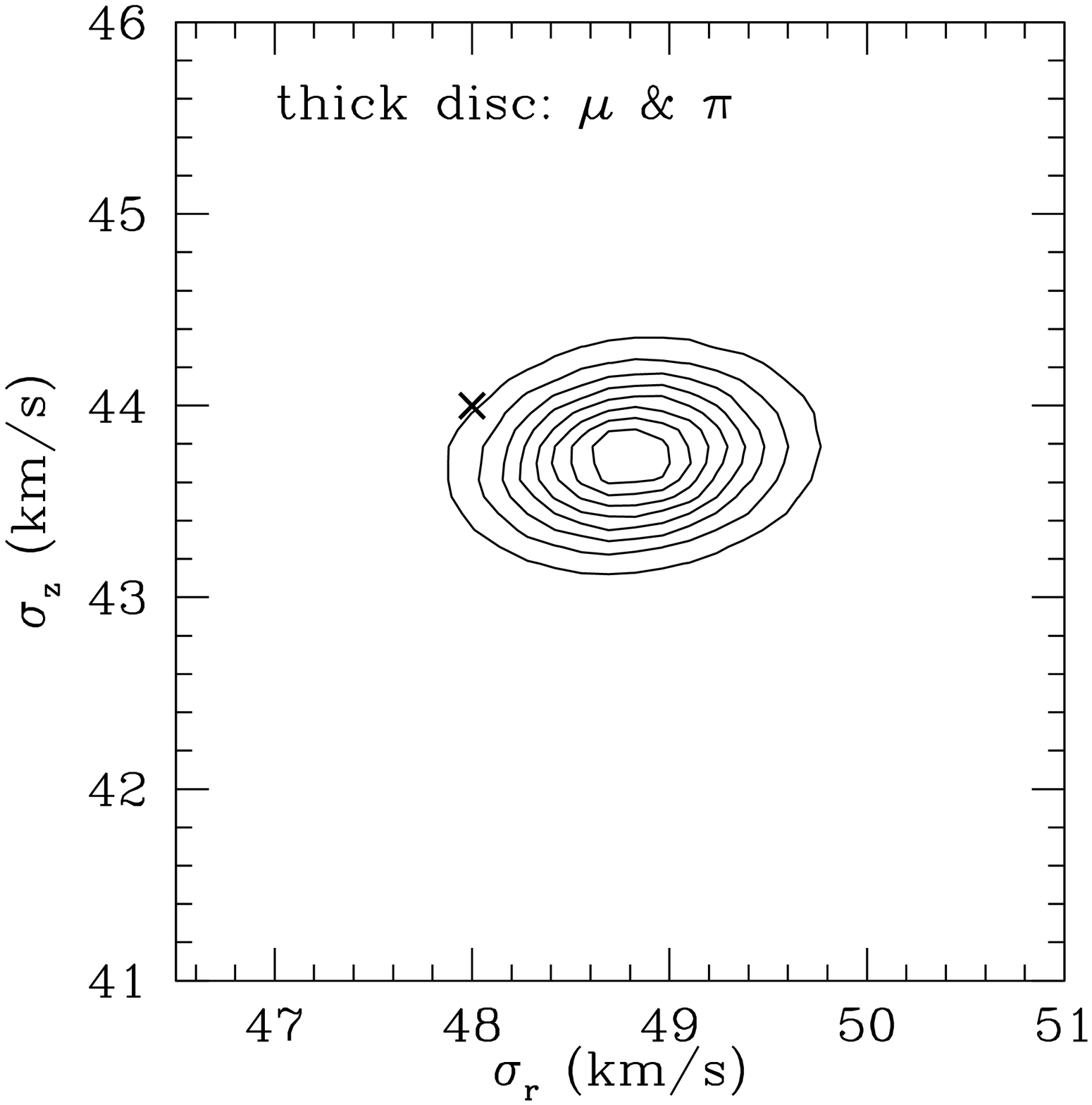}
\includegraphics[width=.3\hsize]{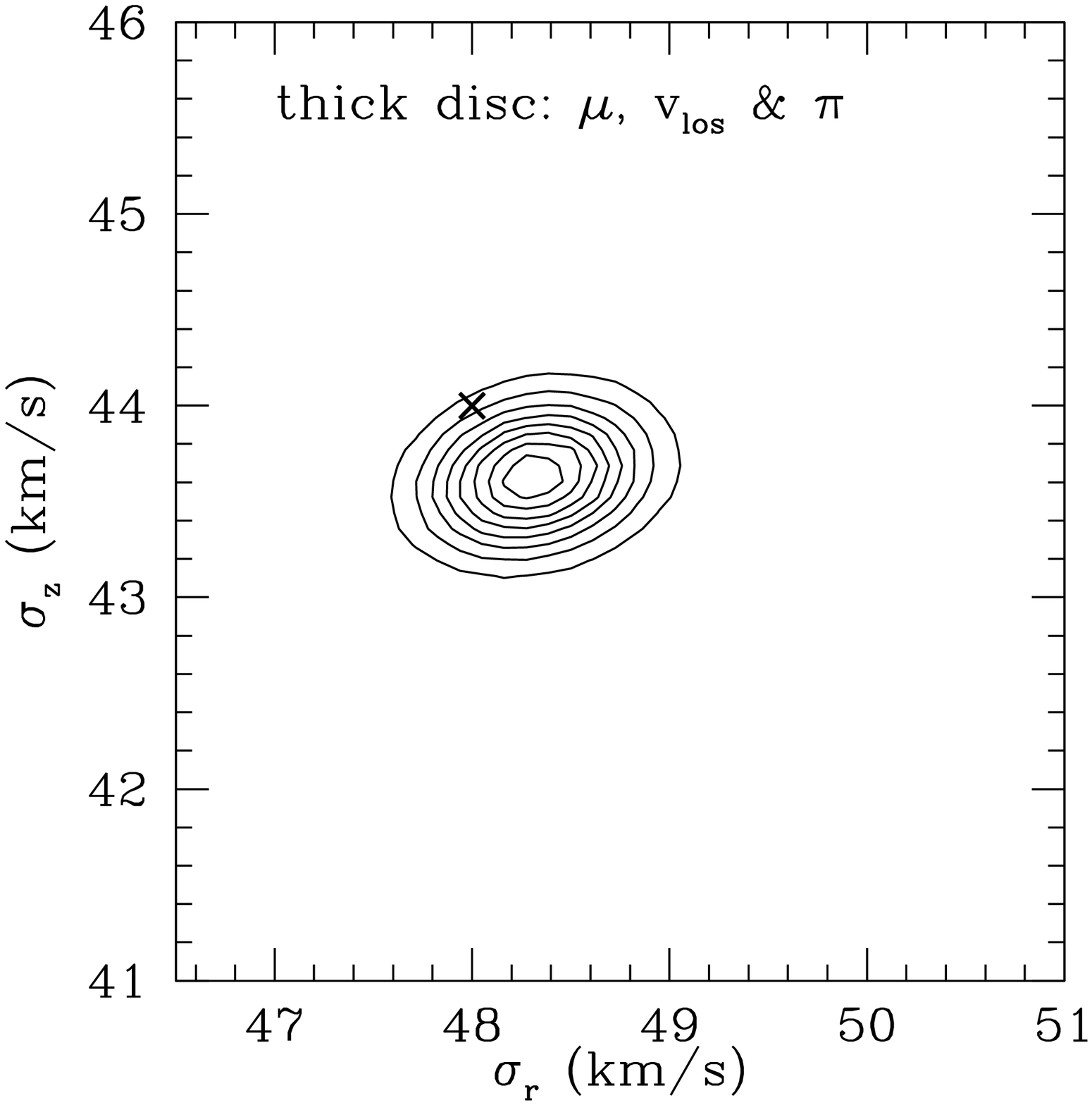}
}
 \caption{As Fig.~\ref{fig:onedisc} but for the case of both thin and thick
discs and using a catalogue with $10\,000$ stars. The top row shows the pdfs
for the parameters of the thin disc after marginalisation over the parameters
of the thick disc, while the bottom row shows those for the thick disc,
similarly marginalised.  }\label{fig:twodisc}
\end{figure*}

Even with error-free data it cannot be possible to constrain the parameters
in the \df\ tightly if the catalogue is small. To determine the size of the
uncertainty that is inherent in using only a finite number of stars we must
do two things: (i) consider a catalogue with error-free data, and (ii)
analyse this catalogue using precisely the tori that were used to generate
the catalogue. This second step eliminates any bias arising from the use of a
finite number of tori to evaluate the integral over $\vJ$ by Monte-Carlo
integration by the following argument. With error-free data, only the star's
true torus will assign a non-zero probability to the star, so it is important
to be sure that this torus is present. Given that it is present, the
Monte-Carlo estimate of the integral over $\vJ$ will be exact, so the only
source of uncertainty in the parameters of the \df\ will be the finite size of
the catalogue.

\figref{fig:noerror} shows the results of running this test on the
single-disc model when we have perfect measurements of proper motion,
parallax and $v_{\rm los}$ for 5\,000 stars (top) or 10\,000 stars (bottom).
The means and dispersions of the parameters in each case are given in the
first two columns of Table~\ref{tab:onedisc}. By resampling the \df\ to
obtain different discrete realisations, one can directly verify that the
formal errors shown in Table~\ref{tab:onedisc} are a fair indication of the
uncertainty in the \df\ that is inherent in the size of the stellar
sample.  The decrease in the uncertainty of both $\sigma_{r0}$ and
$\sigma_{z0}$ with increasing number $N$ of stars in the catalogue is
consistent with $\sigma_{i0}\propto1/\surd N$. With $5\,000$ stars the
irreducible statistical uncertainty is $\sim50$ percent of the uncertainty
with Gaia-like errors in the proper motions, even with no further
information.  The addition of the parallax to the data set reduces the
uncertainty with Gaia-like errors to very little more than the uncertainty
with perfect data.

\subsection{Case of both thin and thick discs}\label{sec:twodisc}

\begin{table}
\caption{Means and dispersions in $\!\kms$ of the parameters of the \df\ recovered from data
made with a Galaxy that has both thin and thick discs, the analysed catalogue
containing 10\,000 stars}
\begin{tabular}{lccc}\label{tab:twodisc}
&$\mu$ only&$\mu,\varpi$&$\mu,\varpi,v_{\rm los}$\\
\hline
$\sigma_{r0}$ &$25.7\pm0.99$&$27.2\pm0.39$&$27.1\pm0.33$\\
$\sigma_{z0}$ &$19.4\pm0.78$&$20.3\pm0.32$&$19.9\pm0.26$\\
\hline
$\sigma_{r0}$ &$48.8\pm0.91$&$48.8\pm0.47$&$48.3\pm0.34$\\
$\sigma_{z0}$ &$43.3\pm0.53$&$43.7\pm0.29$&$43.6\pm0.25$\\
\end{tabular}
\end{table}

\figref{fig:twodisc} and Table~\ref{tab:twodisc} show results for the more
realistic case of two discs with realistic velocity dispersions. In all three
cases the catalogue contained $10\,000$ stars and $100\,000$ tori were used
in the analysis.  

When only proper motions are available, the correlation between $\sigma_{r0}$
and $\sigma_{z0}$ is now rather strong for both thin and thick discs. Adding
parallaxes produces a more dramatic improvement in precision than in the case
of the single-disc model. Adding line-of-sight velocities also produces a
material improvement in accuracy, presumably because the errors of $5\kms$ in
the data are a smaller fraction of typical random velocities than in the case
of the pure thin-disc model. In this test the fraction of the stars in each
disc was fixed at its true value. If this fraction is allowed to vary, the
recovered parameters have larger uncertainties because a trade-off is
possible between the velocity dispersion of the thicker component and the
fraction of the stars it carries: the lower the dispersion, the more stars it
can carry. In the real world this degeneracy must be broken by using chemical
information: the thick disc is fundamentally the component with high
[$\alpha$/Fe].

\section{Discussion}\label{sec:discuss}

In  this paper we have laid out the formalism of an approach to the
interpretation of data from  surveys, and have shown that it works in principle.
However, we have only scratched the surface of the overall problem.
Within the existing framework, we need to:

\begin{itemize}

\item Chart the growth in the uncertainties of the parameters as the
measurement errors of the observations increase.

\item Determine the precision to which the Galactic potential can be
determined by fitting the potential in parallel with the \df, rather than
considering it known.

\end{itemize}

The only barrier to investigating  these questions is  computational cost:
if we change either the measurement errors or the gravitational potential, we
have to re-evaluate the likelihood of the catalogue from scratch rather than
by re-using the saved contributions of each torus to each star. The second of
the questions above is of particular interest because upon its answer depends
the prospects for pinning down the Galaxy's dark-matter density to good
precision.

The computational cost of evaluating the likelihood of a catalogue is
dominated by the integrals of sight-lines in equation (\ref{eq:short}). Since
there is no dependence between these integrals for different stars, the
computation is parallelisable, and we do employ multiple cores.

Two other directions requiring urgent exploration are:
\begin{itemize}
\item Extend the formalism to models that possess more than one stellar
population, each population being characterised by its own \df\ and
distribution in the
colour--luminosity plane.

\item Extend the formalism to include additional observables, such as colour
indices, measures of [Fe/H], [$\alpha$/Fe], etc. 
\end{itemize}

The first of these directions is conceptually straightforward and not
especially computationally costly. It is mandatory in the sense that it would
introduce correlations between kinematics, chemistry and the luminosity
function that were first identified in the Galaxy over half a century ago.
Operationally it is easy: one simply takes the \df\ to be a superposition of
\df s with a different colour--luminosity distribution for each component
\df. The presence of more than one population in the system should sharpen
our ability to constrain the gravitational potential from a given catalogue
as it does in the simpler context of dwarf spheroidal galaxies
\citep{AmoriscoE}.

One could extend the range of observables considered in a trivial way, but at
some point it becomes sensible to widen the scope of the scheme to include
both a model of the chemical evolution of the Galaxy and the strong
constraints on stellar parameters yielded by the theory stellar evolution.
We hope to publish details of such a formalism shortly. The scope of the
exercise is then broadened to diagnosing the history of the Galaxy as well
as its present state, and the resulting constraints on \df\ and gravitational
potential would reflect the information regarding stellar distances that is
normally embodied in ``photometric distances''.

An  issue that must at some point be addressed is:
\begin{itemize}
\item The impact of using an inappropriate form of the \df.
\end{itemize}

 The \df\ of the Galaxy certainly does not consist of a superposition of the
quasi-isothermal \df s we have employed here, and the question arises as so
how accurately the real \df\ can be represented by our functional form (or
some other tractable form). That is, once we have found the best-fitting \df\
of our form, we want to answer the question ``is our catalogue statistically
consistent with being drawn from the chosen \df.'' At an elementary level
this question is readily addressed by drawing pseudo-catalogues from the
model \df\ and for each such catalogue calculating $L$ from equation
(\ref{eq:trial}) -- if the value of $L$ obtained from the real catalogue lies
within the range of those obtained for the pseudo-catalogues, then the model
\df\ is clearly consistent with the data. In reality this exercise will
always reveal that $L$ for the real catalogue lies outside the range defined
by the pseudo-catalogues, because the Galaxy is an extremely complicated
system, full of structure that is not included in the model \df. Before we
can pass judgement on our fitted \df\ we have to understand what it fails to
represent: are there large-scale discrepancies between the predictions of the
\df\ and the observations, or does it merely fail to include small-scale
fluctuations that one is not endeavouring to represent at this level? Given
the limitations on binning the real catalogue that were described in Section
\ref{sec:binning}, addressing this question will be hard.

An issue in this area that can be more easily addressed is ``what impact the
use of an inappropriate form of the \df\ will have on the inferred
gravitational potential and dark-matter density.'' We have to expect that an
inappropriate form of the \df\ will distort our perception of the
gravitational potential because when the \df\ is wrong, we cannot expect the
likelihood of the data to be maximised by the true potential. Clearly an
erroneous potential will yield an erroneous dark-matter distribution. This
question could be rigorously addressed by drawing catalogues from N-body
simulations and comparing the potential and dark-matter densities inferred
from the catalogue with the real ones.

A final topic that merits discussion is
\begin{itemize}
\item Optimising the design of surveys
\end{itemize}
Large resources are currently being invested in surveying the Galaxy, and in
principle it would be useful before committing these resources to determine
the most cost-effective strategy by asking questions such as ``is it more
useful to measure $N$ line-of-sight velocities with errors of $5\kms$ or
$N/2$ velocities with errors of $3\kms$? ``Would it be more useful to
increase by half a magnitude the magnitude limit of a satellite's parallax
measurements, or to measure line-of-sight velocities for a tenth of the
stars? The methods described in this paper make it possible to compute
precise answers to such questions.

In this context an obvious step to take now is to investigate the
degeneracies between the various parameters of Galaxy models, and ask which
combination of data types is most effective at breaking such degeneracies. In
this paper we varied only the velocity scales of the \df. Ideally we would
also vary the other parameters in the \df, such as the scale lengths $R_\d$
and the relative normalisations of the thin and thick discs, and also the
parameters of the gravitational potential, the Sun's velocity, etc. The
extent to which there are degeneracies between these parameters, will depend
on the richness of the data, and it behoves us to understand the relative
power of different data sets before we design a survey.

\section{conclusions}\label{sec:conclude}

It seems inevitable that we will extract the promised science from current
and upcoming surveys of the Galaxy by comparing the surveys' catalogues with
models that have been ``observed'' with biases matched to those of the
surveys. A major goal of the surveys is to map the Galaxy's dark-matter
distribution, which is possible only in so far as the Galaxy can be presumed
to be in a steady state. Consequently equilibrium dynamical models are of
particular importance. 

The number of variables that are commonly measured for each star is large -- in
addition to six phase-space coordinates and the apparent magnitude, one or more
colours, the effective temperature, the surface gravity and two measures of
metallicity are routinely measured because our ambitions to unravel
the star-formation and metal-enrichment histories of the Galaxy turn on the
availability of such data for millions of stars. In Section \ref{sec:binning}
we saw that even if such data were
available for a billion stars, only the crudest estimate of the density of
stars in the high-dimensional data space could be obtained by binning the
data. Therefore it seems inevitable that models will have to be fitted to the
data by maximising the likelihood of the data given a model.

These likelihoods can be calculated only if we have the pdf of the model  in
data space. This requirement is a major difficulty for N-body models, for
which it is in principle no easier to determine the pdf than it is for the
Galaxy. 

For these reasons we believe models based on analytic \df s and orbital tori
have unique potential for the scientific exploitation of large surveys of the
Galaxy. In Section \ref{sec:pseudodata} we described how to produce a discrete
realisation from a torus model, and we have made extensive use of such
realisations in our tests.

In Section \ref{sec:strategy} we developed the formalism required to
determine likelihoods for a torus model under some simplifications. The
principal simplification was the neglect of all variables other than sky
position, space velocities and apparent magnitude. Since in this framework
there is no possibility of distinguishing different types of stars, for
example metal-poor ones, or $\alpha$-enhanced ones or constrain the age of a
star, this level of analysis is only appropriate for a Galaxy that consists
of a single stellar population.  Therefore, this exercise is an artificial
one, but nonetheless a vital one from which we can build towards work of
greater complexity and realism.

The key equation (\ref{eq:trial}) expresses the log-likelihood as a sum of
integrals down the line-of-sight through individual tori. Since a large
number of tori must be used if unbiased results are to be obtained, the
computational cost of these integrals is large. Fortunately, once they have
been evaluated for a given catalogue and gravitational potential, the
likelihood of the data given any \df\ can be quickly computed. Therefore the
optimisation of the \df\ in a given potential is straightforward. 

In Section \ref{sec:test} we tested the algorithm by using it to reconstruct
the \df\ from catalogues containing either $5\,000$ or $10\,000$ stars with
Gaia-like errors. We found that it performed as expected. The irreducible
statistical uncertainty in the \df\ scales roughly inversely with the square
root of the number of stars in the catalogue, and for $5\,000$ stars 
amounts to $\sim0.5$ per cent on the velocity dispersions, which is $\sim50$ per
cent of the uncertainty on the \df\ found with only measurements of the 
proper motion (with Gaia-like uncertainties). 
By adding Gaia-like measurements of
parallax for $5\,000$ stars to those of proper motion, one can drive the
uncertainty in the \df\ almost down to the irreducible statistical
uncertainty.

The uncertainties quoted above are remarkably small given the small sizes of
the catalogues analysed. The uncertainties will grow, probably significantly,
when the potential is varied alongside the \df. They will also grow with the
sophistication of the \df, although one may hope that this growth will be to
a large extent countered by enrichment of the data to include
spectrophotometric data that must accompany any attempt to decompose the
Galaxy into several populations with distinct \df s.

The uncertainties on parameters that we recover include the effects of
measurement errors and statistical uncertainty arising from the finite number
of stars in the analysed catalogue, but do not make allowance for the use of
only a finite number of tori. An insufficient supply of tori will lead to
biases in the results. The entropy of the probability distribution defined by
equation (\ref{eq:defspk}) being small is a sure sign that too few tori are
being used, but the  surest check that enough tori are being used
is  to draw a fresh sample of tori from the final \df\ and to re-determine
the pdf of the \df's parameters using these tori. If the pdf differs
materially from the original one, more tori are required.

It is remarkable that the \df\ can be recovered from proper motions alone
because with the broad luminosity function used here the data by themselves
contain negligible distance information. The distance information that is
required to pin down the \df\ and thus establish the typical velocities of
stars is provided by the gravitational potential, which sets a scale-height
at any velocity dispersion, and by the solar motion through the logic of
traditional secular parallaxes. Since only limited distance information is
available when the data are restricted to proper motions, the two velocity
scales of the \df, $\sigma_{r0}$ and $\sigma_{z0}$ have correlated errors.

Adding parallaxes to the data set eliminates this correlation as well
as diminishing the scale of the uncertainties. Further adding line-of-sight
velocities with uncertainties of $5\kms$ has at most a small impact on the
results.

In Section \ref{sec:discuss} we discussed several directions for further
work. One urgent step is to extend the formalism to include the spectral
characteristics of stars, such as metallicity, colour, effective temperature
and surface gravity. The main issue here is how best to extend our models to
predict their distributions. There is more than one way that this can be
done, so we reserve this topic for a future paper. A related issue is how
best to combine constraints from different surveys, for example an
astrometric survey such as Gaia with a spectroscopic one such as the ESO-Gaia
survey. One option is to analyse the catalogues one after the other, using
constraints obtained from the first analysis to define the prior used in the
second analysis. Alternatively it may be possible to refine the definition of
the selection function $\phi(\vJ)$ in such a way that catalogues with
different selection criteria
can be analysed simultaneously.

In this paper we have neglected extinction. Given a three-dimensional model
of the interstellar medium, it would be straightforward if computationally
costly, to allow for reddening and extinction. Ideally, the model of the ISM
would be refined in parallel with the Galaxy model, but it is not yet
apparent how this could be accomplished in practice.

A key topic is the extent to which the Galaxy's gravitational potential
can be constrained by various bodies of data. In principle this is a trivial
extension of the present work, but it is computationally expensive. 

Computational cost is a real issue. The cost of analysing a catalogue is
proportional to the number of its stars times the mean number of tori
$\overline{n}_*$ that might make a non-negligible contribution to each star's
probability. Although the total number of tori needed increases with the
precision of the data, $\overline{n}_*$ should be roughly constant between
catalogues, so computational cost should scale with the number of stars in
the catalogue.  However, this situation may be improved by binning stars by 
position on the sky, and evaluating the appropriate line-of-sight integral 
(eq. \ref{eq:short}) simultaneously for all the stars in a bin for a given
torus, under the approximation that their position on the sky is that of 
the centre of the bin. Approaches like this which reduce computational 
effort are likely to prove essential in scaling up this algorithm from 
working with the pseudo-catalogues of $10\,000$ stars used here to analysing 
surveys such as RAVE, with $\lta500\,000$ stars observed and, looking further 
ahead, the $\sim$$10^9$ stars in the Gaia catalogue.

\section*{Appendix:  Proof that $L$ is stationary for the true DF}

By the Monte-Carlo integration theorem, the sum over $\alpha$ in the
definition (\ref{eq:trial}) of the likelihood $L$ can be replaced by an
integral over $\vu$ times the pdf of the stars in that space, which is
$\pr(\vu|f_{\rm t})$, where $f_{\rm t}$ is the true \df. Hence
 \begin{equation}
L=\int\d^7\vu\,\pr(\vu|f_{\rm t})\ln[\pr(\vu|f)],
 \end{equation}
 so
\begin{equation}
{\p L\over\p a}=\int\d^7\vu\,\pr(\vu|f_{\rm t}){\p\pr(\vu|f)/\p a
\over\pr(\vu|f)}.
\end{equation}
 When $f=f_{\rm t}$ the denominator cancels with the pdf on top and we find
\begin{equation}
{\p L\over\p a}\bigg|_{f=f_{\rm t}}=\int\d^7\vu\,{\p\pr(\vu|f_{\rm t})\over\p a}.
\end{equation}
 This vanishes because it is the derivative of $\int\d^7\vu\,\pr(\vu|f_{\rm
 t})=1$.

\label{lastpage}
\end{document}